\documentclass[12pt,njp]{iopart}
\usepackage{iopams}  

\expandafter\let\csname equation*\endcsname\relax
\expandafter\let\csname endequation*\endcsname\relax

\usepackage{graphicx,amssymb,amsmath,epsfig,color,enumerate,bm,subfig,multirow,algorithm,algorithmic}

\epsfclipon

\usepackage[T1]{fontenc}
\usepackage{citesort}

\begin{document}
 \title{Master equation approach to the stochastic accumulation dynamics of bacterial cell cycle}
 \author{Liang Luo$^{1,2,\dagger}$, Yang Bai$^{3\dagger}$ and Xiongfei Fu$^{3}$}
\address{$^1$ Department of Physics, Huazhong Agricultural University, Wuhan 430070, China}
\address{$^2$ Institute of Applied Physics, Huazhong Agricultural University, Wuhan 430070, China}
\address{$^3$ CAS Key Laboratory for Quantitative Engineering Biology, Guangdong Provincial Key Laboratory of Synthetic Genomics, Shenzhen Institute of Synthetic Biology, Shenzhen Institutes of Advanced Technology, Chinese Academy of Sciences, Shenzhen, 518055, China}
\address{$^{\dagger}$ These authors contributed equally}
\ead{luoliang@mail.hzau.edu.cn}

\begin{abstract}

The mechanism of bacterial cell size control has been a mystery for decades, which involves the well-coordinated growth and division in the cell cycle. The revolutionary modern techniques of microfluidics and the advanced live imaging analysis techniques allow long term observations and high-throughput analysis of bacterial growth on single cell level, promoting a new wave of quantitative investigations on this puzzle. Taking the opportunity, this theoretical study aims to clarify the stochastic nature of bacterial cell size control under the assumption of the accumulation mechanism, which is favoured by recent experiments on species of bacteria. Via the master equation approach with properly chosen boundary conditions, the distributions concerned in cell size control are estimated and are confirmed by experiments. In this analysis, the inter-generation Green's function is analytically evaluated  as the key to bridge two kinds of statistics used in batch-culture and mother machine experiments. This framework allows us to quantify the noise level in growth and accumulation according to experimental data. As a consequence of non-Gaussian noises of the added sizes, the non-equilibrium nature of bacterial cell size homeostasis is predicted, of which the biological meaning requires further investigation. 

\end{abstract}

\maketitle

\section{Introduction}

Bacterial cells manage to grow and divide into two cells in a cell cycle. As a common observation, the sizes of the bacterial cells varies from one to another, but the fluctuation is well controlled\cite{wagner14,biswas14pnas,suckjoon15}. The fascinating problem arises how bacteria coordinate the mass growth and division processes to keep the cell size stable. 
As a first approach to investigate the cell size control mechanism, the experiments were performed to measure the mean cell mass in various growth conditions. In 1958, Schaechter, Maal\o{}e, and Kjeldgaard discovered the relation between the mean cell size and the growth rate in their experiments with extraordinary precision\cite{schaechter58}, which set the foundation of studies on cell size control for 60 years. Following their work, there had been a wave of fruitful investigations in bacterial physiology for quantitative understanding of the cell size distribution in an exponentially growing culture\cite{maclean61,collins62,koch62,powell64,koch66}. The control mechanism beneath the distribution was, however, not clear due to limited information provided by experiments in that age. In the recent decade, the newly developed mother machine experiment makes it possible to track cell cycles of single cells for long term\cite{suckjoon10}, which initiated a new wave of studies on cell size control\cite{suckjoon10,wagner14,suckjoon15,elf16,wagner19,suckjoon19}. With the help of the single-cell data, deterministic models of bacterial cell size control have been established on the level ofmean values\cite{suckjoon15,willis17,suckjoon18}. By incorporating randomness in these models, attempts for understanding the ever-fluctuating cell cycles were also made in various approaches\cite{suckjoon15,singh16,marco17,biswas17,biswas18,suckjoon19,willis20,nieto20,jia21,jia21prx}.

As the key technique boosting the recent wave of studies, mother machine refers to a new bacterial culturing method realized by the microfluidic techniques, where channels of micrometre-width are fabricated to trap single bacterium while continuously supplying it with fresh medium\cite{suckjoon10}. In the experiments, only one mother cell is constantly trapped in the channel while all the daughter cells are pushed out during the cell growth and replication. One can easily measure the birth size, division size, and inter-division time of mother cells for generations in such setup. It was observed that the added size of the cell between successive divisions is uncorrelated with the birth size. The ``adder principle'' was hence established as a common strategy of bacterial cell size control, which suggests that a cell attempts to add a constant size before the next division\cite{wagner14,suckjoon15}. The following question appears how a cell could sense the added size. 
A natural guess comes that the added size is sensed by certain circuits of chemical reactions. The reaction locales of the circuits are, however, non-trivial. They can not be uniformly distributed in the bulk of cytoplasm, since the chemical reactions in bulk is relevant to the concentration but not the absolute abundance, where the total cell volume plays no role. 
In bacteria with no nucleus, the reaction locales may either be on DNA or cell membrane. 
A likely candidate of the reaction is the assembly of the contractile ring on the membrane\cite{suckjoon19,wu12}. It was reported in the recent experiment the assembly is re-initiated in each newborn cells and the division can be conducted only if the assembly is completed\cite{suckjoon19}. Based on the above knowledge, an accumulation model was introduced that a newborn cell accumulates certain molecules as the adder indicator and it divides when the absolute abundance of the indicator reaches a threshold\cite{suckjoon15,baiyang20}. 

Quantitative modelling of the cell growth and division processes dates back to the pre-DNA age\cite{rahn32,kendall48}, where cell growth was considered as simple and clean stochastic processes due to the lack of detailed biological knowledge. Enlightened by the success of the microbiologists in the 1960s, Bell, Anderson and their colleagues developed the mathematical framework in the fashion of master equation for systematic description of cell growth and division processes\cite{bell67,bell67-2,bell68,bell69,bell71}. In the new era of bacterial physiology,  inter-generation dynamics\cite{amir14,you15,marco17,thomas18,jafarpour19} has been intensively studied aiming to reveal the cell control mechanism from the single-cell data provided by mother machine experiments. The models integrating the dynamics at both single cell level and the population level were also developed in the context of modern biological knowledge\cite{suckjoon15,biswas14prl,biswas14pnas,singh16,biswas18,willis20,nieto20,jia21}. While stochastic features were inevitably incorporated in the models, a clear investigation on the stochastic accumulation model is still absent. 

In this theoretical study, we investigate a transparent model for the stochastic growth-accumulation-division process with only two free parameters. The distributions concerned in cell size control are estimated, which are well confirmed by the experiments. As the central result of this work, we demonstrate the inter-generation Green's function is the key to bridge two kinds of statistics employed in the batch-culture experiments and that of the mother machine ones. It is analytically shown that not only the mean value but also the distribution of added size is independent of the birth size, which is elucidated in the picture of subordinated random walk. The analysis on experimental data shows that the noise levels significantly vary for different growth conditions. It hence suggests that in addition to growth rate, the noise strengths are necessary parameters to describe the bacterial cell cycle on the level of distributions. Evaluating the entropy production of the inter-generation stochastic dynamics, the non-equilibrium nature of bacterial cell size homeostasis is further predicted.




The paper is organized as follows. Section \ref{sec_model} introduces the stochastic accumulation model in the framework of master equation. Section \ref{sec_bc} revisits the cell size distribution of the exponentially growing culture. Section \ref{sec_mm} studies the inter-generation stochastic dynamics derived from the comprehensive model, where the analytic expression of the distribution of added size is shown. Section \ref{sec_noise} estimates the noise levels of the experiments. Section \ref{sec_ne} reveals the non-equilibrium nature of the cell size control mechanism. Section \ref{sec_disc} contains discussions, including brief historical remarks, the connections with the other models, the differences and connections between the statistics in batch culture and mother machine, the noise levels in various growth conditions, and some biological implications. Section \ref{sec_cc} gives a summary. 

\section{Cell cycle in the framework of master equation}
\label{sec_model}

\subsection{The deterministic population dynamics}

\begin{figure}[tbp]
\centering
\includegraphics[width=.7\textwidth]{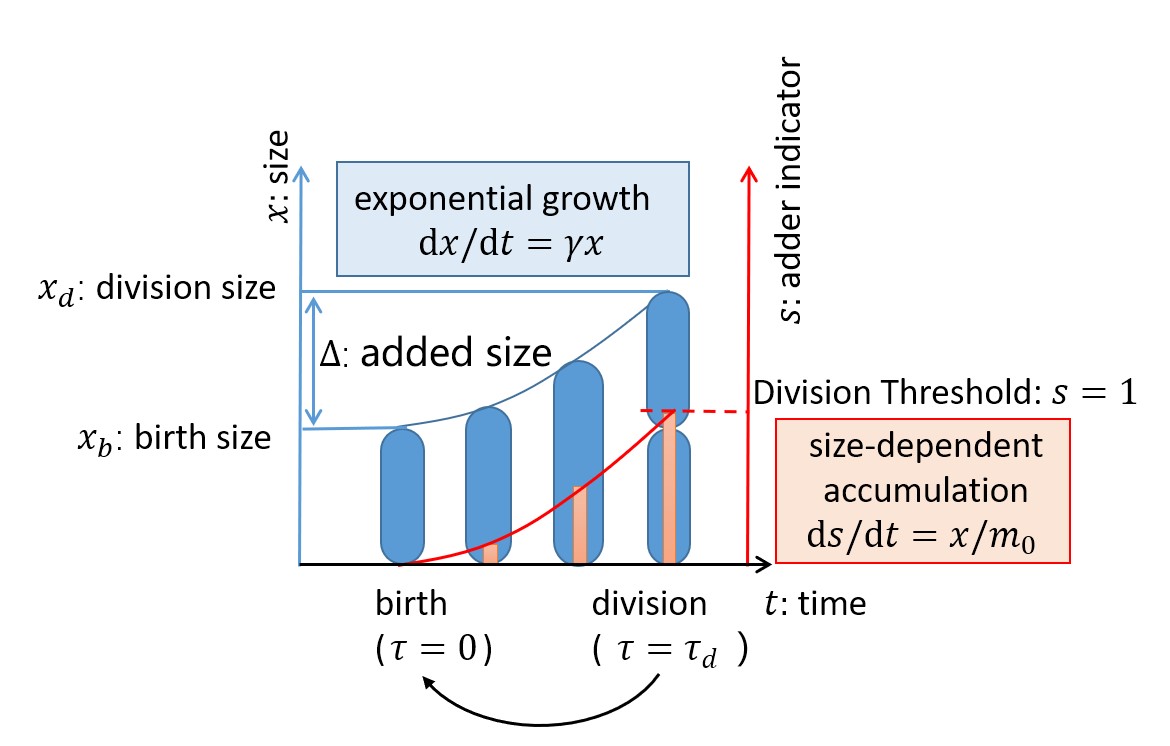}
\caption{\label{fig_dm}The deterministic growth-accumulation-division model. 
}
\end{figure}

This work studies the accumulation model, which is illustrated in Fig.\ref{fig_dm}. According the model, the cell size $x$ exponentially grows with the rate $dx/dt=\gamma x$, while it accumulates certain adder indicator $s$ from the newborn value $s=0$ with the rate $ds/dt=x/m_0$. The cell division happens when the adder indicator reaches the division threshold $s=1$. Assuming even division, the birth sizes of the daughter cells are half of the division size of the mother, $x_b=x_d/2$. 

The framework of master equation of the population dynamics is employed to investigate the model. Let us denote the number of the cells of size $x$, age $\tau$ and the adder indicator $s$ at time $t$ in the whole population by $u(x,s,\tau, t)$. Taking cell age, size growth and indicator accumulation into account, the time course of $u$ is governed by the master equation 
\begin{equation}
\label{eq_cme}
\frac{\partial }{\partial t}u=-\frac{\partial }{\partial \tau}u-\frac{\partial }{\partial x}(\gamma x u)-\frac{\partial }{\partial s}(\frac{x}{m_0} u),
\end{equation}
where we have adopted the exponential cell growth assumption that the rate coefficients are set as  $k^{(x)}=\gamma x$ with the cell growth rate $\gamma$ and $k^{(s)}=x/m_0$. We will see later $m_0$ sets the mean added size for each cell cycle. 

The accumulation model assumes cell accumulates certain protein $s$ as the adder indicator from $s=0$ at birth. The cell divides at the threshold $s_0=1$. The division current, i.e. the number of division events in unit time,  can be written as
\begin{equation}
J_{d}(x)=\int_0^{\infty}d\tau\;\frac{x}{m_0}\left.\frac{\partial}{\partial s}u(x,s,\tau)\right\vert_{s=1}.
\end{equation}
To avoid additional complexity, we restrict ourselves in the case of perfect even-division, of which the two daughter cells have the same size. It suggests the birth current $J_{b}(x)=2J_{d}(2x)$. 
Noting that the age and the adder indicator are reset for the newborn cells, the full division-birth process can be written as the boundary condition
\begin{equation}
\label{eq_cbc}
\left.\frac{\partial}{\partial t}u(x,s,\tau,t)\right\vert_{s=0,\tau=0}=2\int_0^{\infty}d\tau\;\frac{2x}{m_0}\left.\frac{\partial}{\partial s}u(2x,s,\tau)\right\vert_{s=1}.
\end{equation}
Eq.(\ref{eq_cme}) and Eq.(\ref{eq_cbc}) define the population dynamics of exponentially growing bacteria colony. 

In this model, the growth rate $\gamma$ is the only parameter relevant to the time scale. 
The equation is thus invariant under rescaling by $x'=x/(\gamma m_0)$, $\tau'=\gamma\tau$, $t'=\gamma t$, which gives
\begin{equation}
\label{eq_cme1}
\frac{\partial }{\partial t'}u=-\frac{\partial }{\partial \tau'}u-\frac{\partial }{\partial x'}(x' u)-\frac{\partial }{\partial s} (x'u).
\end{equation}
The dynamics of the rescaled cell size is independent of the growth rate $\gamma$ and the typical size $m_0$. The famous bacteria growth law observed by Schaechter-Maal\o{}e-Kjeldgaard\cite{schaechter58}, $\left<x\right>\propto\gamma m_0$, is hence a direct consequence of the rescaling. The rescaling approach remains valid for the slow growth cases, of which the recent experimental study suggested the correction on $m_0$ by $\gamma m_0\propto(\gamma +\gamma_0)$\cite{baiyang20}. 

The dynamics defined by the above continuous master equation is deterministic. To be explicit, let us consider a colony starting from a cell of known birth size $x_0$, with the initial condition $u(x,s,\tau)\vert_{t=0}=\delta(x-x_0)\delta(s)\delta(\tau)$. According to Eq.(\ref{eq_cme}) and Eq.(\ref{eq_cbc}), the joint distribution $u$ would ever be of zero-width. It is obviously not the case of a real bacterial colony. The stochastic theory including noises in cell growth and division is hence required. 

\subsection{The stochastic population dynamics}

Master equation is naturally a language for Markov processes on discrete states. The deterministic dynamics defined above can be easily generalized to the stochastic process on discrete cell sizes $\{x_i=i\delta_x\}$ and adder indicators $\{s_j=j\delta_s\}$, which follows the master equation
\begin{equation}
\label{eq_dme0}
\frac{\partial}{\partial t}u_{ij}=-\frac{\partial}{\partial\tau}u_{ij}-\frac{\gamma}{\delta_x}\left(x_i u_{ij}-x_{i-1}u_{(i-1)j}\right)-\frac{1}{m_0\delta_s}\left(x_i u_{ij}-x_i u_{i(j-1)}\right),
\end{equation}
where $u_{ij}$ denotes $u(x_i,s_j,\tau,t)$ for simplicity.
Cell division is defined by the boundary condition
\begin{equation}
\label{eq_dbc0}
\frac{\partial}{\partial t}\left.u_{i0}\right\vert_{\tau=0}=\frac{2}{m_0\delta_s}\int_0^{\infty}d\tau\; x_{2i}u_{2i,n_0},
\end{equation}
where $n_0=1/\delta_s$ is the division threshold for the adder indicator. 
Acknowledging the proteins are generally synthesized in the cell in a bursty manner\cite{xie06}, $n_0$ would be understood as the number of bursts required for synthesizing adequate specific division-related protein. 

Similar to that in the continuous case, scaling can be introduced by $x'_i=x_i/(\gamma m_0)$, $t'=\gamma t$, $\tau'=\gamma\tau$, $s'=s$, $\delta'_x=\delta_x/(\gamma m_0)$ and $\delta'_s=\delta_s$. 
The rescaled discrete master equation is written as
\begin{equation}
\label{eq_dme}
\frac{\partial}{\partial t'}u_{ij}=-\frac{\partial}{\partial\tau'}u_{ij}-\frac{1}{\delta'_x}\left(x'_i u_{ij}-x'_{i-1}u_{(i-1)j}\right)-\frac{1}{\delta'_s}\left(x'_i u_{ij}-x'_i u_{i(j-1)}\right),
\end{equation}
while the boundary condition is given by
\begin{equation}
\label{eq_dbc}
\frac{\partial}{\partial t'}\left.u_{i0}\right\vert_{\tau'=0}=\frac{2}{\delta_s}\int_0^{\infty}d\tau'\; x'_{2i}u_{2i,n_0}
\end{equation}
Compared with the continuous case, there are two additional parameters left in the rescaled dynamics, $\delta_x$ and $\delta_s$, which come from the noises in growth and division. In some cases, it would be convenient to introduce the inversed noise strength $\alpha=1/\delta'_x$ and $\beta=1/\delta'_s$. In the rest of the paper, the dynamics and distributions are investigated in the rescaled form, where $\gamma$ and $m_0$ are set to $1$ and the noise parameters play the key roles.  

\begin{figure}[tbp]
\centering
\includegraphics[width=.7\textwidth]{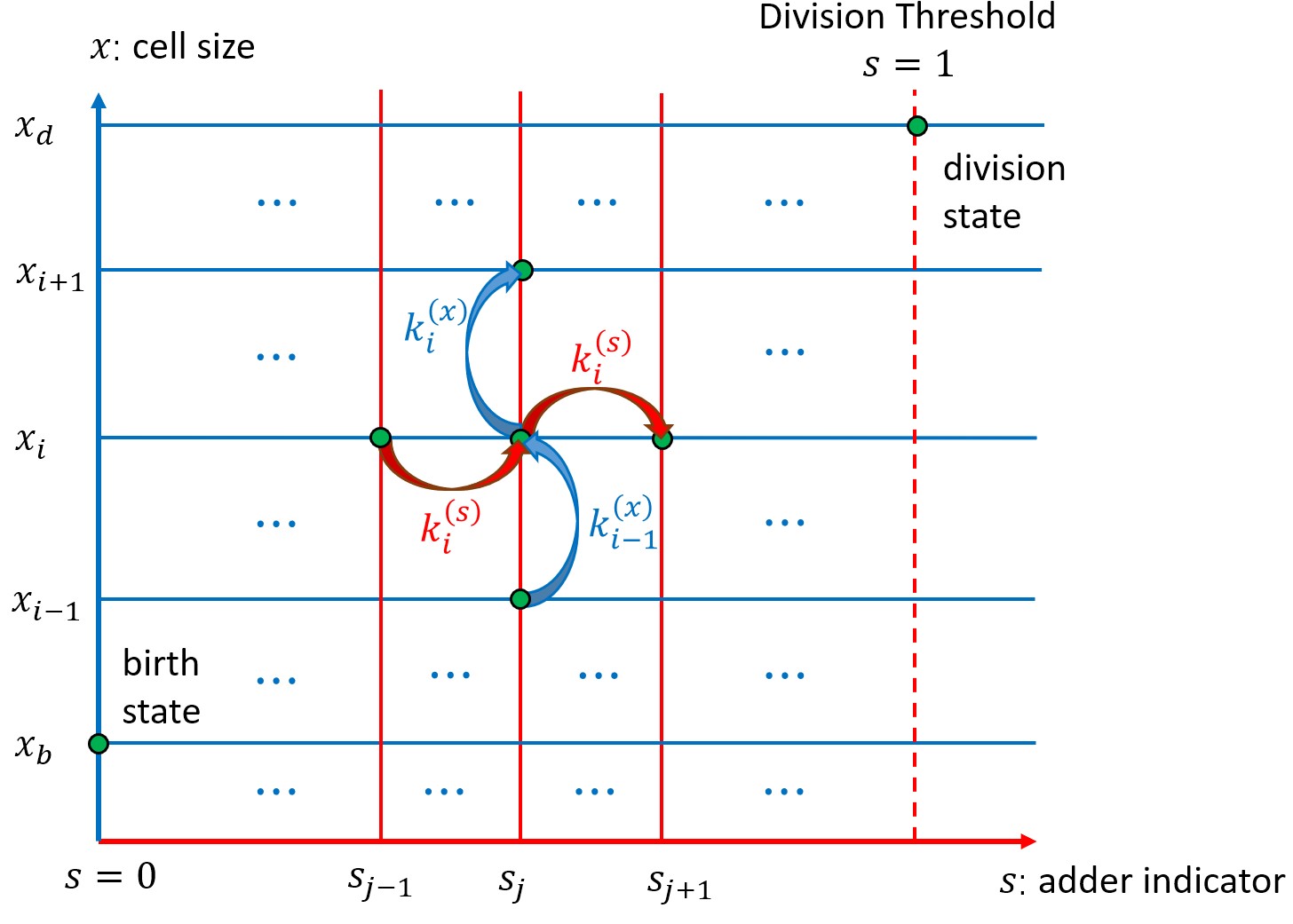}
\caption{\label{fig_sm}The stochastic growth-accumulation-division process as random walk on two-dimensional lattic of cell sizes $\{x_i=i\delta_x\}$ and adder indicator $\{s_j=j\delta_s\}$. The walk starts from the birth state $(x_b,s=0)$. At state $(x_i,s_j)$, the walk either jumps to $(x_{i+1,s_j})$ with the rate $k^{(x)}_i=\alpha x_i$ or to $(x_i,s_{j+1})$ with the rate $k^{(s)}_i=\beta x_i$. Cell division happens when the adder indicator reaches the threshold $s=1$ at the division state $(x_d,s=1)$. 
}
\end{figure}

The stochastic processes described by the above master equation can be understood as random walk on the two-dimensional lattice $\{x_i,s_j\}$, as shown in Fig.\ref{fig_sm}. The random walk initiates at the cell birth with $(x=x_b,s=0)$. For each state $(x_i,s_j)$, the walker jumps either to $(x_{i+1},s_j)$ by the rate $k^{(x)}_i=\alpha x_i$ or to $(x_i,s_{j+1})$ by the rate $k^{(s)}_i=\beta x_i$, until the cell division at $(x_d,s=1)$. The pseudocode for simulation in Gillespie's fashion\cite{gillespie77} is provided in \ref{app_pseudocode} for better understanding, through most of the results are obtained by analytic and numerical approaches instead of simulation.

Since the model is defined for discrete $\{x_i, s_j\}$ and continuous $\tau$, we would like to clarify that the expressions are often simplified as $P(x_i,s_j,\tau)=(d\tau)^{-1}\text{Prob}\left(x_i,s_j,\tau\in(\tau,\tau+d\tau)\right)$, with the capital letters. The probability density functions (PDF) are introduced for comparisons among various $\delta_x$, $\delta_s$, and the experimental data, which are denoted by the lower letters as $p(x,s,\tau)=(\delta_x\delta_s)^{-1}P(x_i,s_j,\tau)$. 
There have been already series of studies on bacterial cell cycle, where the stochastic dynamics were constructed in different ways\cite{suckjoon15,biswas14prl,singh16,warsawpre19,nieto20,jia21,kendall48,amir14,you15,marco17,biswas17}. The connections and differences among the previous models and our model will be discussed in Sec. \ref{sec_comp}.

\begin{table}
\centering
\caption{\label{tab_notation} Notation List} 
\begin{indented}
\lineup
\item[]\begin{tabular}{@{}*{7}{l}}
\br                              
\0\0& Notation & Definition &\cr 
\mr
\multirow{9}{2cm}{Deterministic model} & $u(x,s,\tau,t)$ (or $u$ for short) & number of cells of specified $x$, $s$, $\tau$ at time $t$\cr
& $x$& cell size\cr 
& $s$  & adder indicator \cr
& $\tau$ & cell age \cr
& $t$ & time\cr
& $\gamma$ & growth rate \cr
& $m_0$ & constant that sets the mean added size \cr
& $k^{(x)}=\gamma x$ & the increment rate of cell size\cr
& $k^{(s)}=x/m_0$ & the increment rate of adder indicator\cr
\hline
\multirow{6}{2cm}{Rescaled variables and parameters} & $x'=x/(\gamma m_0)$ & the rescaled cell size\cr
& $s'=s$ & the rescaled adder indicator\cr
& $\tau'=\gamma\tau$ & the rescaled cell age\cr
& $t'=\gamma t$ & the rescaled time\cr
& $k^{(x)}=x'$ & the rescaled increment rate of cell size\cr
& $k^{(s)}=x'$ & the rescaled increment rate of adder indicator\cr
\hline
\multirow{4}{2cm}{Stochastic model} & $u_{ij}(\tau,t)$ (or $u_{ij}$ for short) & number of cells of specified $x_i$, $s_j$, $\tau$ at time $t$\cr
& $x_i=i\cdot\delta_x$ & the discrete cell size\cr
& $s_j=j\cdot\delta_s$ & the discrete adder indicator\cr
& $\delta_x$, $\delta_s$ & minimal increments of cell size and adder indicator\cr
\hline
\multirow{4}{2cm}{Rescaled noise parameters} & $\delta'_x=\delta_x/(\gamma m_0)$, $\delta'_s=\delta_s$ & the rescaled minimal increments\cr
& $\alpha=1/\delta'_x$, $\beta=1/\delta'_s$ & inversed noise strengths\cr
& $k_i^{(x)}=x_i/\delta'_x=\alpha x_i$ & the rescaled increment rate for cell size\cr
& $k_i^{(s)}=x_i/\delta'_s=\beta x_i$ & the rescaled increment rate for adder indicator\cr
\hline
\multirow{4}{2cm}{Cell cycle} & $x_b$ & birth size\cr
& $x_d$ & division size\cr
& $\Delta=x_d-x_b$ & added size\cr
& $\tau_d$ & inter-division time\cr
\br
\end{tabular}
\end{indented}
\end{table}

\section{Batch culture and the cell size distribution}
\label{sec_bc}

In biological studies, bacteria are usually batch cultured where they are kept in the exponentially growing phase by frequent transfer between batches of fresh medium before it exhausts the environment nutrient. Due to least disturbance on bacterial cells and the ability to keep a steady growth rate, the batch culture procedure was widely employed in the studies of the 1950s and 1960s when the cell size distribution has been intensively studied. Due to the rapidly increasing cell number in the crowding colony, the batch culture experiment is not convenient for single-cell tracking even with the help of modern techniques. The statistics in this case hence usually involve all the cells in different stages, from the newborn one to the ready-to-division one. 

The master equation Eq.(\ref{eq_dme}) with the boundary condition Eq.(\ref{eq_dbc}) provides a straightforward description of the exponentially growing population. One can numerically integrate the dynamics over time and observing the development of a bacteria colony. 
The cell number of the colony can be evaluated by summing over all the states as
\begin{equation}
\label{eq_ncell}
N=\sum_{i,j}\int d\tau\;u_{ij}.
\end{equation}
As Eq.(\ref{eq_dme}) defines the growth process where the cell number is conserved, cell doubling arises only during the division process defined by Eq.(\ref{eq_dbc}). Considering a colony starting from one cell, the doubling events behave as steps of $N(t)$ in the initial generations. The synchronization blurs afterwards due to noises. In the long time limit, a steady state of exponential growth is achieved, where the cell number grows smoothly. 

\begin{figure}[tbp]
\centering
\includegraphics[width=.7\textwidth]{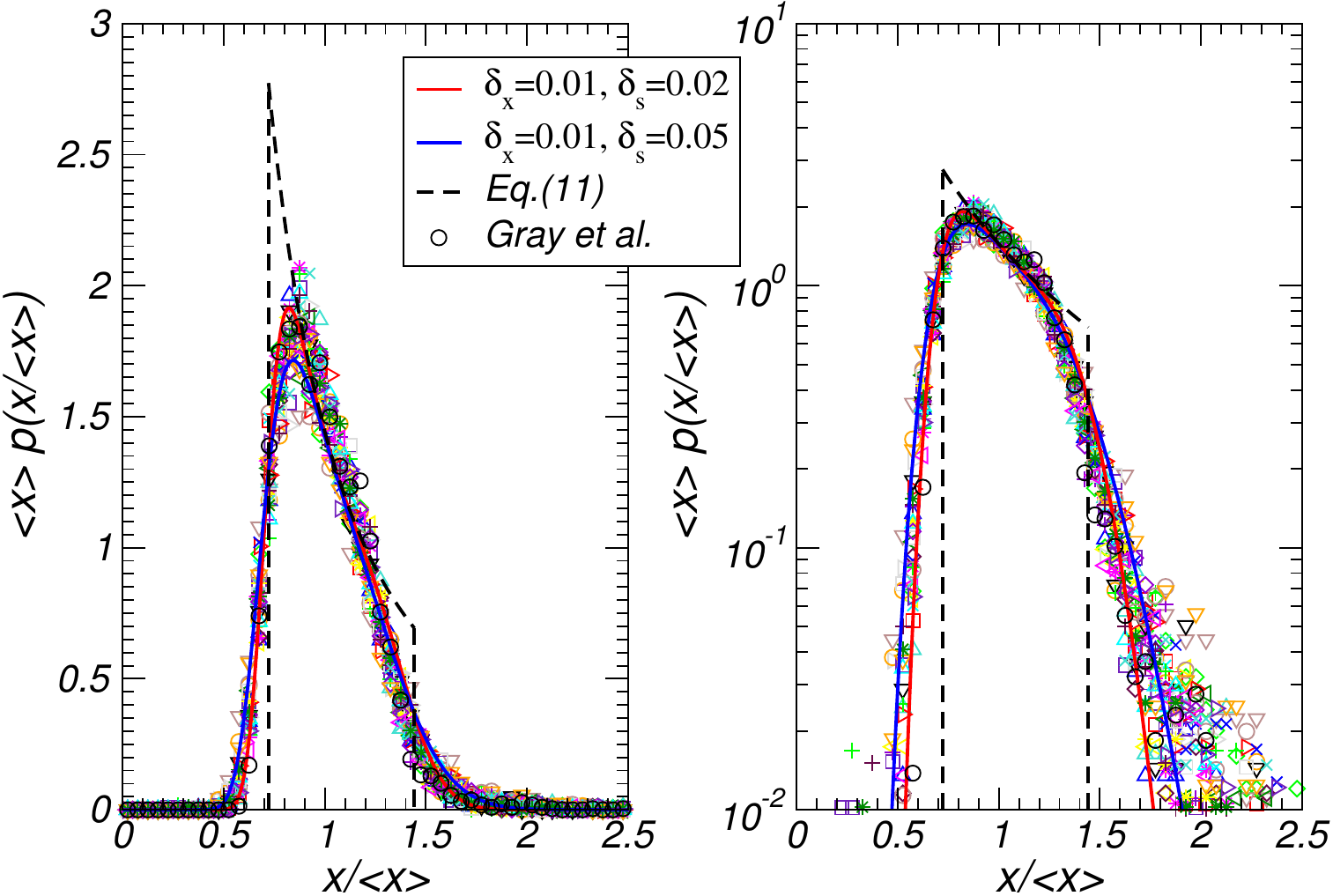}
\caption{\label{fig_xdist}  The probability density function of the cell size  of the exponentially growing population, scaled by the mean value. (Left) The solid lines are the steady distributions obtained by the master equation. The dashed line shows the inverse square law of cell size distribution (Eq.(\ref{eq_isl})). The symbols are from the experiments by Gray et al.\cite{wagner19}. (Right) Same as left, but in log-scale. } 
\end{figure}

In the current framework, the cell size distribution can be evaluated as the marginal probability
\begin{equation}
P(x_i)=\frac{1}{N}\sum_j\int d\tau\;u_{ij}. 
\end{equation}
The corresponding PDF is defined as $p(x)\equiv P(x_i)/\delta_x$. In the long time limit, the steady distribution can be achieved, which is shown in Fig.\ref{fig_xdist}. To our surprise, the cell size distribution is so well confirmed by the data from the experiment\cite{wagner19}. Revisiting the historical literatures, we realize that the microbiologists in the 1960s have already obtained similar results in a series of fruitful investigations of different approaches other than master equation. (A relevant historical remark is included in Sec. \ref{sec_hr}.) 
It is known since then the inverse-square law for the exponentially growing and evenly divided cells,
\begin{equation}
\label{eq_isl}
p_{ss}(x)=\frac{2x_b}{x^2}, \text{ for }x_b<x<x_d,
\end{equation}
which is shaped by the growing cells with size $x$ ranging from the birth size $x_{b}$ to the division size $x_{d}=2x_{b}$\cite{maclean61,koch62}. Including the fluctuation of birth size and division size in consideration, the distribution is blurred with fast-decaying tails, as shown by the solid lines in Fig.\ref{fig_xdist}. Similar results were also noticed by Koch\cite{koch66}. 
While the experiments in the 1960s have roughly confirmed the results, the high-quality modern experiment\cite{wagner19} provides more details of the distribution. As shown in Fig.\ref{fig_xdist}, the experiment suggests a rather fat tail for large cell size, which may indicate postponed cell divisions due to certain unknown biological reasons. 

\section{Mother machine and the inter-generation stochastic dynamics}
\label{sec_mm}

Being developed in the 21st century, the mother machine technique provides a new way to cultivate bacteria. Since only one of the two daughters of each division is tracked, the cell number in the channels of mother machine is constant over generations. Noting this, the modified boundary condition is introduced for mother machine as
\begin{equation}
\label{eq_dbc1}
\frac{\partial}{\partial t}\left.u_{i0}\right\vert_{\tau=0}=\frac{1}{\delta_s}\int_0^{\infty}d\tau\; x_{2i}u_{2i,n_0},
\end{equation}
where the prefactor for the number of tracked newborn cell is changed to $1$. One can of course investigate the cell size distribution of the population, similar to Sec.\ref{sec_bc}. It is, however, more interesting to investigate the distributions of birth sizes $x_b$, division sizes $x_d$ and inter-division times $\tau_d$ of the mother cells as they can be recorded as series over generations through the mother machine experiment, which makes it ideal to study the cell size control mechanism.

\subsection{Master equation and Green's function of the inter-generation stochastic dynamics}
\label{sec_gf}

The cell size control mechanism is usually formulated as the generation-wise dynamics of the division size (or birth size) from the $i$th generation to the $(i+1)$th one as
\begin{equation}
\label{eq_dld}
x_d^{(i+1)}=\frac{1}{2}x_d^{(i)}+\Delta, 
\end{equation}
where perfect even-division is assumed for the birth size $x_b^{(i+1)}=x_d^{(i)}/2$. The added size can be decomposed as $\Delta=\Delta_0+\xi(x_d^{(i)})$, where $\Delta_0$ is the expected added size and $\xi$ is the noise. It is referred to the discrete-time Langevin dynamics in literature, where the noise $\xi$ has been assumed following normal distribution\cite{you15} or log-normal distribution\cite{amir14,marco17} based on arguments. 

The inter-generation stochastic dynamics (Eq.(\ref{eq_dld})) can be equivalently described by another master equation, which concerns the evolution of division size distribution over generations as
\begin{equation}
\label{eq_dldme0}
P^{(i+1)}(x_d)=\int dx'\; G(x_d\vert x'_d) P^{(i)}(x'_d),
\end{equation}
where Green's function $G\left(x_d^{(i+1)}\vert x_d^{(i)}\right)$ is the conditioned probability of the division size $x^{(i+1)}_d$ for a cell with the given birth size $x_b^{(i+1)}=x_d^{(i)}/2$.
In the case of discrete cell size, the above equation is written as
\begin{equation}
\label{eq_dldme}
P^{(i+1)}_m=\sum_n G_{mn}P^{(i)}_n,
\end{equation}
where $P^{(i)}_n=P^{(i)}(x_d=x_n)$ and $G_{mn}=G\left(x^{(i+1)}_d=x_m\vert x^{(i)}_d=x_n\right)$.
In the more compact matrix form, it is given as
\begin{equation}
\label{eq_dldmatrix}
P^{(i+1)}=\hat{G}P^{(i)}.
\end{equation}
It shows that the inter-generation stochastic dynamics are fully determined by Green's function, which works as a matrix on the vector of division size distribution. 

\begin{figure*}[tbp]
\centering
\includegraphics[width=.8\textwidth]{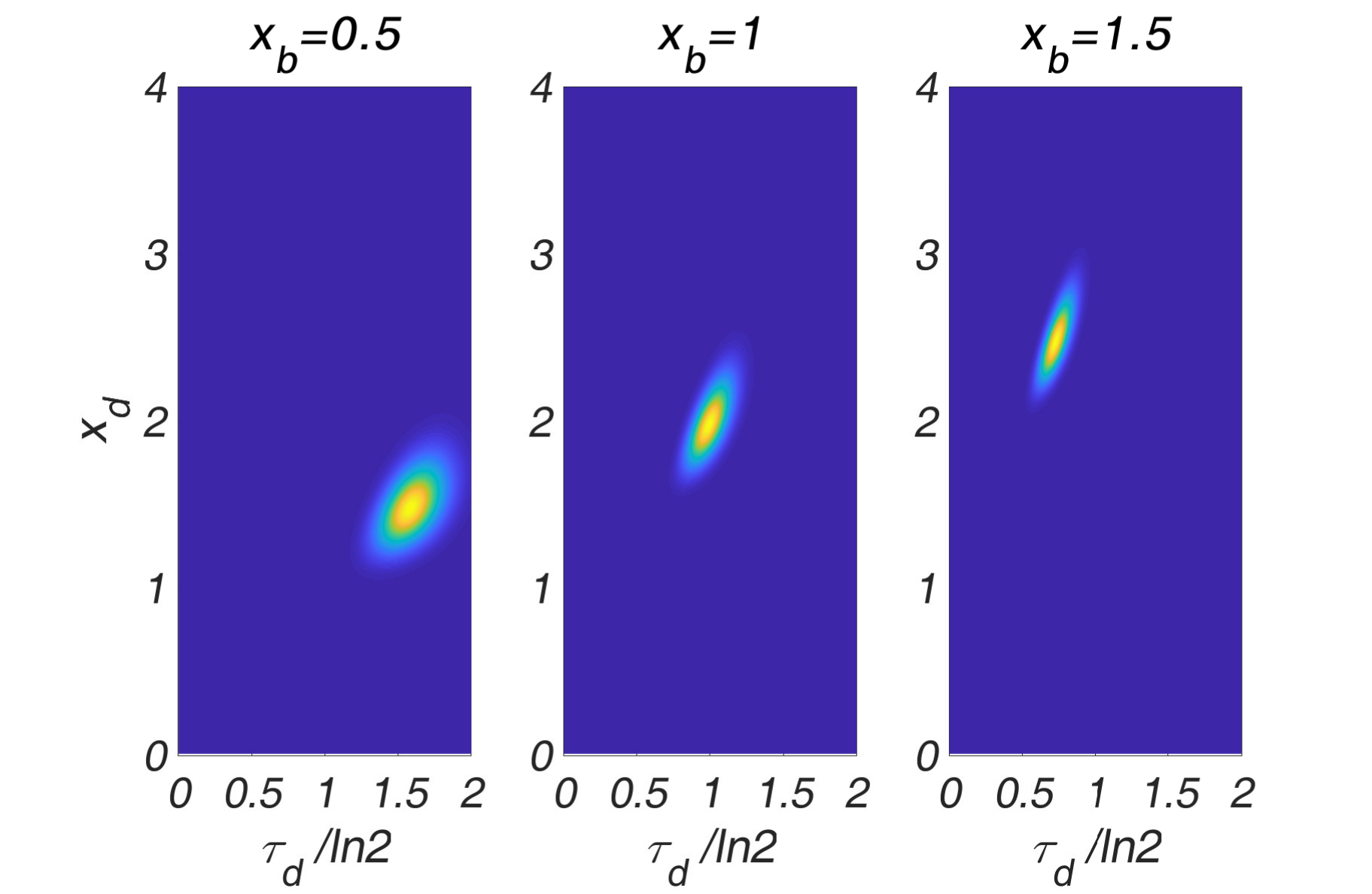}
\caption{\label{fig_fp_heatmap} The heatmap of the first-passage distribution $F(x_d,\tau_d\vert x_b)$ for $\delta_x=0.01$, $\delta_s=0.02$ and various $x_b$. }
\end{figure*}

\subsection{The distribution of added size}
\label{sec_as}

In general, the inter-generation dynamics is a consequence of the intra-generation dynamics, i.e. the growth and division processes. Assuming the current stochastic accumulation model, the Green's function can be evaluated from the one-generation processes of a cell starting with a given birth size $x_b=x_d^{(i)}/2$ and ending at certain division size $x_d=x_d^{(i+1)}$ when the adder indicator $s$ reaches the threshold. The initial condition can be written as $u\vert_{t=0}=\delta(x-x_b)\delta(s)$. 
Concerning the division size and inter-division time, the full Green's function is given by the first-passage distribution at $s=1$ as
\begin{equation}
\tilde{G}(x_d^{(i+1)},\tau\vert x_d^{(i)})=F\left(x_d=x_d^{(i+1)},\tau_d=\tau\vert x_b=x_d^{(i)}/2\right). 
\end{equation}
Figure \ref{fig_fp_heatmap} shows $F(x_d,\tau_d\vert x_b)$ for three typical $x_b$ in which one may notice the inter-division time decreases with the birth size and the division size increases with it. 

In the generation-wise dynamics described in Sec.\ref{sec_gf}, one focuses on the series of division sizes (or equivalently birth sizes). By integrating over the inter-division time, the Green's function can be obtained as
\begin{equation}
G\left(x_d^{(i+1)}\vert x_d^{(i)}\right)=\int d\tau\;\tilde{G}\left(x_d^{(i+1)},\tau\vert x_d^{(i)}\right). 
\end{equation}
Bearing in mind the adder principle, one may reorganize the Green's function of the successive division sizes as the distribution of the added size by
\begin{equation}
G\left(x_d^{(i+1)}\vert x_d^{(i)}\right)=P\left(\Delta=x_d^{(i+1)}-x_d^{(i)}/2\left\vert x_b=x_d^{(i)}/2\right.\right).
\end{equation}

\begin{figure}[tbp]
\centering
\includegraphics[width=.7\textwidth]{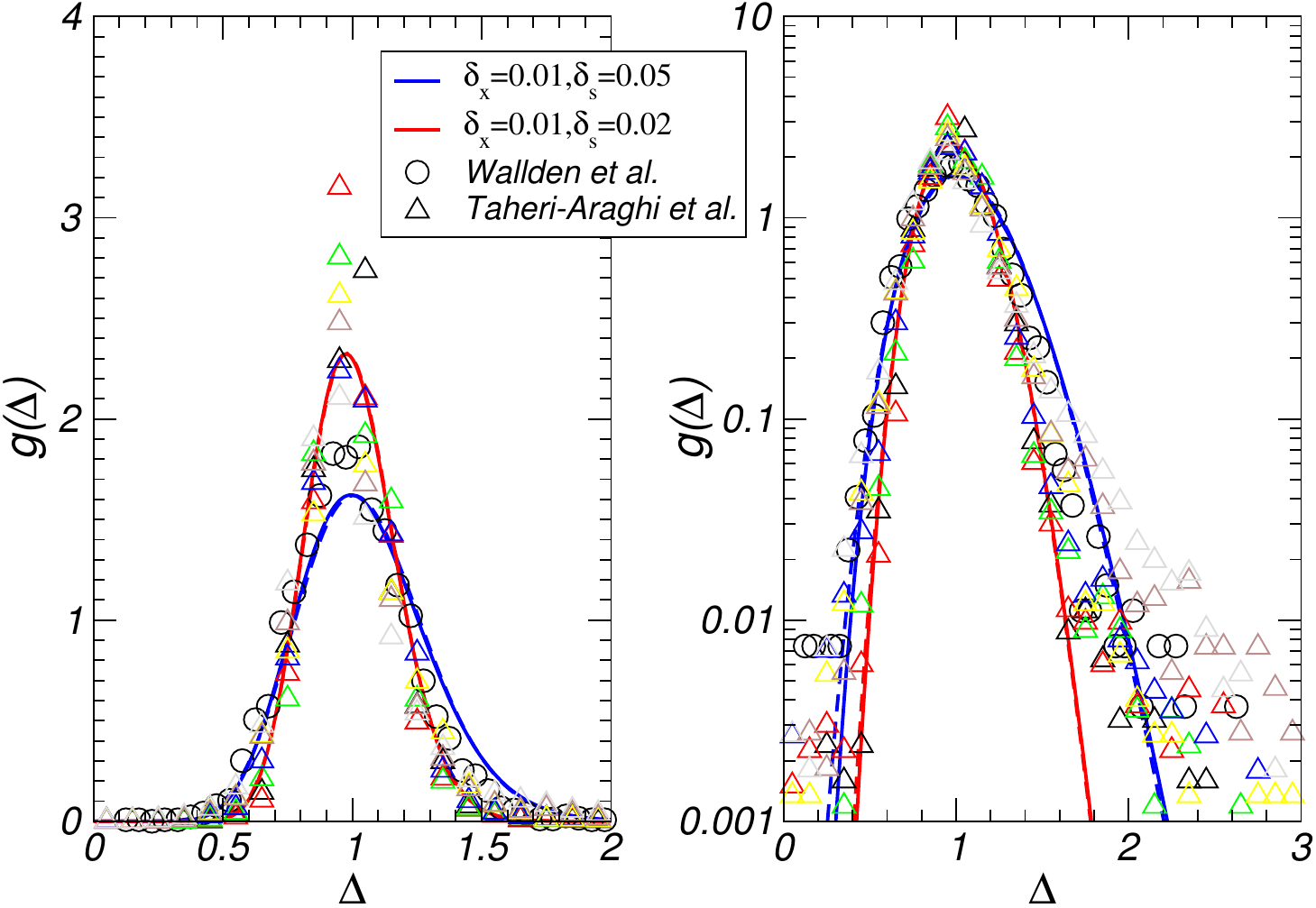}
\caption{\label{fig_gf} The probability density function of the added size $\Delta$. (Left) The solid lines show $g(\Delta)$ according to Eq.(\ref{eq_das}) and Eq.(\ref{eq_das1}), for two typical noise levels. The dashed lines show the log-normal distribution (Eq.(\ref{eq_lognormal})) with the same mean value and variance of distribution given by Eq.(\ref{eq_das1}). The solid lines and dashed lines are well matched, which can be merely distinguished in the tail regions. The symbols show the experimental data of various growth conditions from \cite{suckjoon15,elf16}. (Right) Same as left, but in log-scale. 
}
\end{figure}

It is well known in the deterministic accumulation model that the mean added size depends only on the growth rate but not on the birth size. The current stochastic model provides the opportunity to further investigate the distribution of the added size. As one of the main results of this study, the distribution is explicitly obtained as
\begin{equation}
\label{eq_das}
P(\Delta\vert x_b)=\binom{\alpha\Delta+\beta}{\beta}\frac{\alpha^{\alpha\Delta}\beta^{\beta+1}}{(\alpha+\beta)^{\alpha\Delta+\beta+1}},
\end{equation}
where the inversed noise strength $\alpha=1/\delta_x$ and $\beta=1/\delta_s$ are used for simplicity. 
The corresponding PDF follows:
\begin{equation}
\label{eq_das1}
g(\Delta)=\delta_x^{-1}P(\Delta\vert x_b).
\end{equation}
It can be obtained by directly solving the discrete master equation, of which the technical details can be found in \ref{app_dist}. Another quick and illustrative approach will be soon shown in this section, which explains why Eq.(\ref{eq_das}) is in the form of the negative binomial distribution.

The noise strengths determine the shape of the rescaled distribution of added size, which can be characterized by the mean value, coefficient of variation, and skewness, as shown in \ref{app_noise}. Figure \ref{fig_gf} shows the distributions along with the experimental data\cite{suckjoon15,elf16}, where the theoretical results are presented by lines for illustration without fine-tuning on the parameters. One can read significant differences among the distributions of experiments in various growth conditions, even after rescaling over the mean added size (or saying, rescaling over the growth rate). 
It would be a consequence of different noise levels in various growth conditions, which leads the investigation in Sec. \ref{sec_noise}. 
 
\begin{figure}[tbp]
\centering
\includegraphics[width=.7\textwidth]{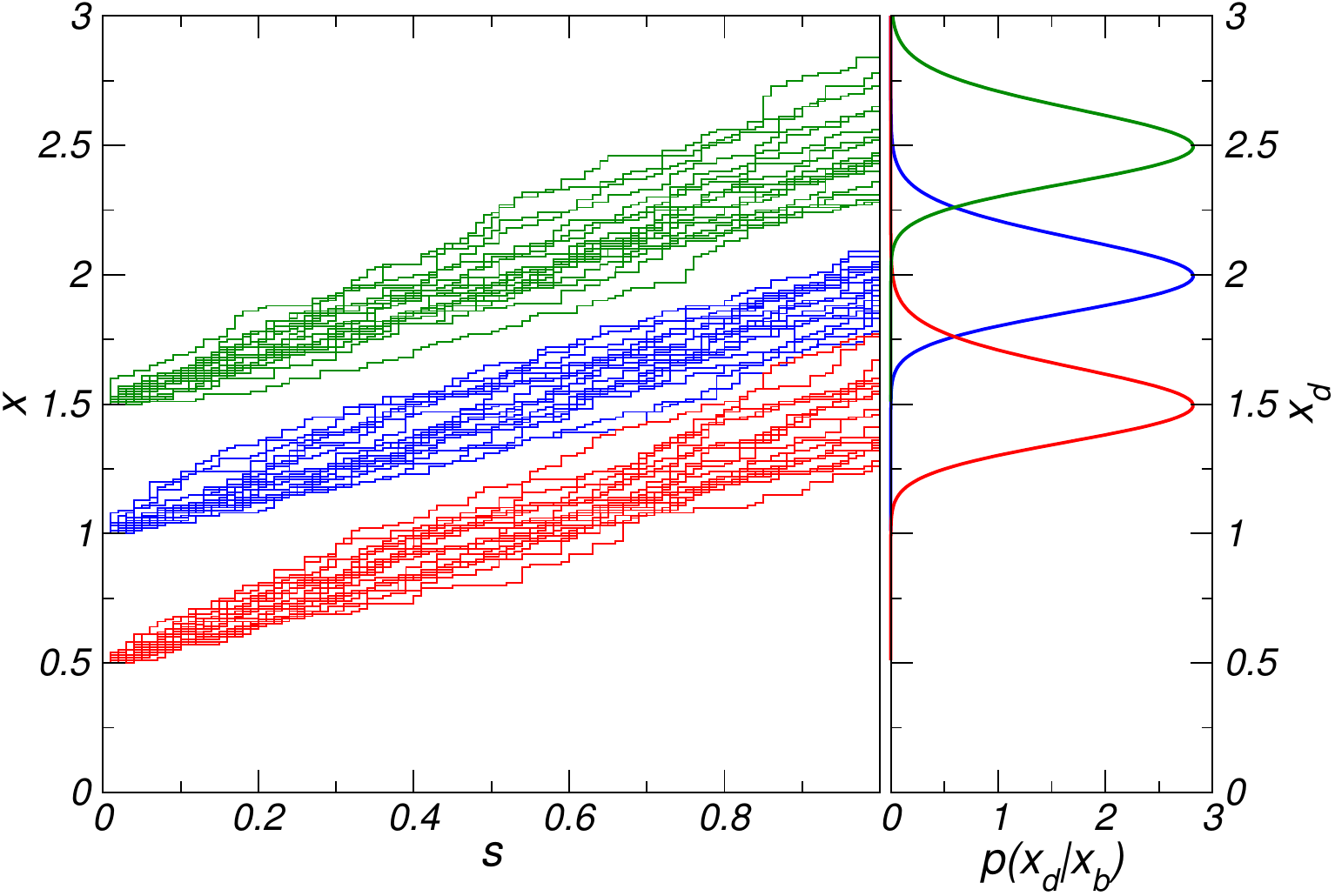}
\caption{\label{fig_rw} (Left) The growth process as random walk on the lattice of $\{x_i,s_j\}$ for three birth sizes, $x_b=0.5$, $1$, and $1.5$, shown in different colors. (Right) The corresponding division size distribution as a shift of Eq.(\ref{eq_das1}), $p(x_d=x\vert x_b)=p(\Delta=x-x_b\vert x_b)=g(x-x_b)$. 
}
\end{figure}

Let's keep in the track of theoretical investigation of the model in this section. 
One may note that $P(\Delta\vert x_b)$ is independent of $x_b$. 
It would be a general feature of the accumulation mechanism, which has also been reported in the experiment\cite{suckjoon15} and a previous theoretical study on a model neglecting growth fluctuation\cite{singh16}. The picture of random walk on the two-dimensional lattice $\{x_i,s_j\}$, as illustrated by Fig.\ref{fig_sm}, would be helpful to clarify it. Since the transition rates depend only on the current state, it is a kind of trap dynamics\cite{haus87}, of which the subordination idea works\cite{bochnerbook,sokolov05,luo14}. Following this idea, the time-dependent process $\{x_i,s_j,t\}$ can be described as the jumps of a random walk on the lattice $\{x_i,s_j,n\}$ subordinated by the random waiting time $\{t_n\}$ for jumps. It leads to Gillespie's algorithm\cite{gillespie77}, of which the trajectory on the lattice and the series of waiting time can be generated independently. Estimating the distribution of division size, only the former one, i.e. $\{x_i,s_j,n\}$, is concerned. It depends on the ratio $r_w=k^{(x)}/k^{(s)}$ but is irrelevant of the scale of the waiting time $ \tau_w=\left(k^{(x)}+k^{(s)}\right)^{-1}$. Noting that $r_w=\alpha/\beta$ is constant for all the states, one can immediately see the random walk $\{x_i,s_j,n\}$ is independent of the initial cell size $x_b$. The first-passage distributions of $x$ at $s=1$ hence have the same shape, as illustrated in Fig. \ref{fig_rw}. The above reasoning can be extended to more general cases, as long as the growth rate $k^{(x)}$ and accumulation rate $k^{(s)}$ depends only on the current cell state while the ratio $k^{(x)}/k^{(s)}$ is independent of cell state. This requirements can be satisfied by most adder dynamics, which explains the collapse of distributions observed in the experiment and the previous model study\cite{suckjoon15,singh16}. More discussion on the biological implication can be found in Sec.\ref{sec_bi}.

A quick and illustrative approach to Eq.(\ref{eq_das}) follows the above insights is to map the random walk to ``Bernoulli trials'' with success probability $q=\beta/(\alpha+\beta)$. The cell divides when $n_s=1/\delta_s+1$ success trials, i.e. $n_s$-steps of accumulation of the adder indicator, are achieved. In this case, the number of failed trials $n_x$, i.e. the steps of cell growth, follows the negative binomial distribution as
\begin{equation}
P(n_x)=\binom{n_s+n_x-1}{n_x}q^{n_s}(1-q)^{n_x}. 
\end{equation}
Noting $\Delta=n_x\delta_x$, $\alpha=1/\delta_x$, $\beta=1/\delta_s$ and the symmetry of combination number $\binom{n}{m}=\binom{n}{n-m}$, one can obtain Eq.(\ref{eq_das}). 

There have been recently a series of investigations\cite{singh16,nieto20,jia21,jia21prx} on a special case of the current model with $\delta_x\rightarrow 0$ (or $\alpha\rightarrow\infty$). In this case, the exponential growth is assumed in the continuous and deterministic manner while cell evolves along serial stages with random waiting times. At the final stage, the division happens. In the case of $\alpha\rightarrow\infty$, the added size distribution given by Eq.(\ref{eq_das}) and Eq.(\ref{eq_das1}) becomes the Erlang distribution with the PDF:
\begin{equation}
p(\Delta)=\frac{\beta^{\beta+1}}{\beta!}\Delta^{\beta}e^{-\beta\Delta}.
\end{equation}
To obtain the above result, Stirling's formula, $\ln n!=n\ln n- n + O(\ln n)$, is used.


The above special case leads to a long argued assumption that the division size follows the log-normal distribution, see e.g.\cite{koch66jtb,amir14,suckjoon15,marco17}. This assumption also follows the insight of the cell cycle as serial stages while the cell grows exponentially with a constant rate. In the case that the number of the stages is large, the inter-division time can be approximated by the normal distribution with the probability density
\begin{equation}
p(\tau_d=t\vert x_b)\simeq\frac{1}{\sqrt{2\pi\sigma_t^2}}\exp\left[-\frac{t-\left<t\right>}{2\sigma_t^2}\right].
\end{equation}
Noting the exponential growth with the constant rate $\gamma$ leads to $x_d=x_be^{\gamma \tau}$, and also $p(\tau_d)d\tau_d=p(x_d)dx_d$, one can obtain
\begin{equation}
p(x_d=x\vert x_b)=\frac{1}{\sqrt{2\pi\sigma_x^2}}x^{-1}\exp\left[-\frac{\left(\ln x-\ln x_b-\gamma\left<t\right>\right)^2}{2\sigma_x^2}\right],
\end{equation}
where $\sigma_x=\gamma\sigma_t$.
Noting $p(\Delta\vert x_b)=p(x_d=\Delta+x_b\vert x_b)$, one can expect the added size follows a shifted log-normal distribution by the probability density
\begin{equation}
\label{eq_lognormal}
p(\Delta\vert x_b)=\frac{1}{\sqrt{2\pi\sigma_x^2}}(\Delta+x_b)^{-1}\exp\left[-\frac{\left(\ln (\Delta+x_b)-\ln x_b-\gamma\left<t\right>\right)^2}{2\sigma_x^2}\right].
\end{equation}
Although the above argument can not justify the independence of added size on $x_b$, it is still supported by our model that Eq.(\ref{eq_lognormal}) with $x_b=\Delta_0$ is a good approximation of $g(\Delta)$, as shown by the lines in Fig.\ref{fig_gf}, of which the deviance can be distinguished merely in the tail regions. 

\subsection{The steady distributions in the long time limit}
\label{sec_ds}

\begin{figure}[tbp]
\centering
\includegraphics[width=.7\textwidth]{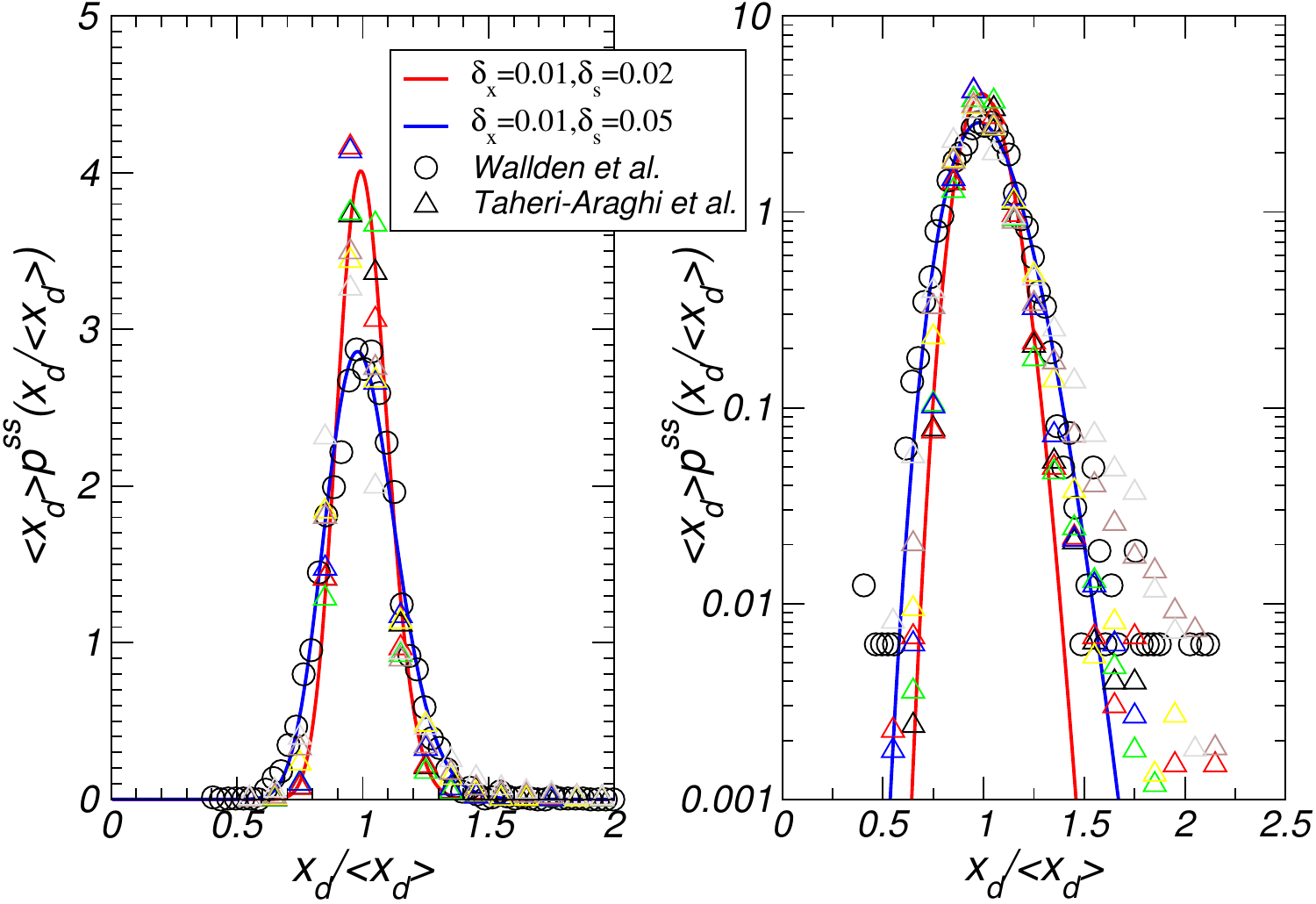}
\caption{\label{fig_xd} The probability density function of the steady distribution of division size normalized by mean value. (Left) The lines show $\left<x_d\right>p^{ss}(x_d/\left<x_d\right>)$ for two typical noise levels. The symbols show the experimental data from \cite{suckjoon15,elf16}. (Right) Same as left, but in log-scale. 
}
\end{figure}

It has been a common practice to measure the steady distributions of division size and inter-division time in long time mother machine experiments. We estimate the steady distributions here.

\begin{figure}[tbp]
\centering
\includegraphics[width=.7\textwidth]{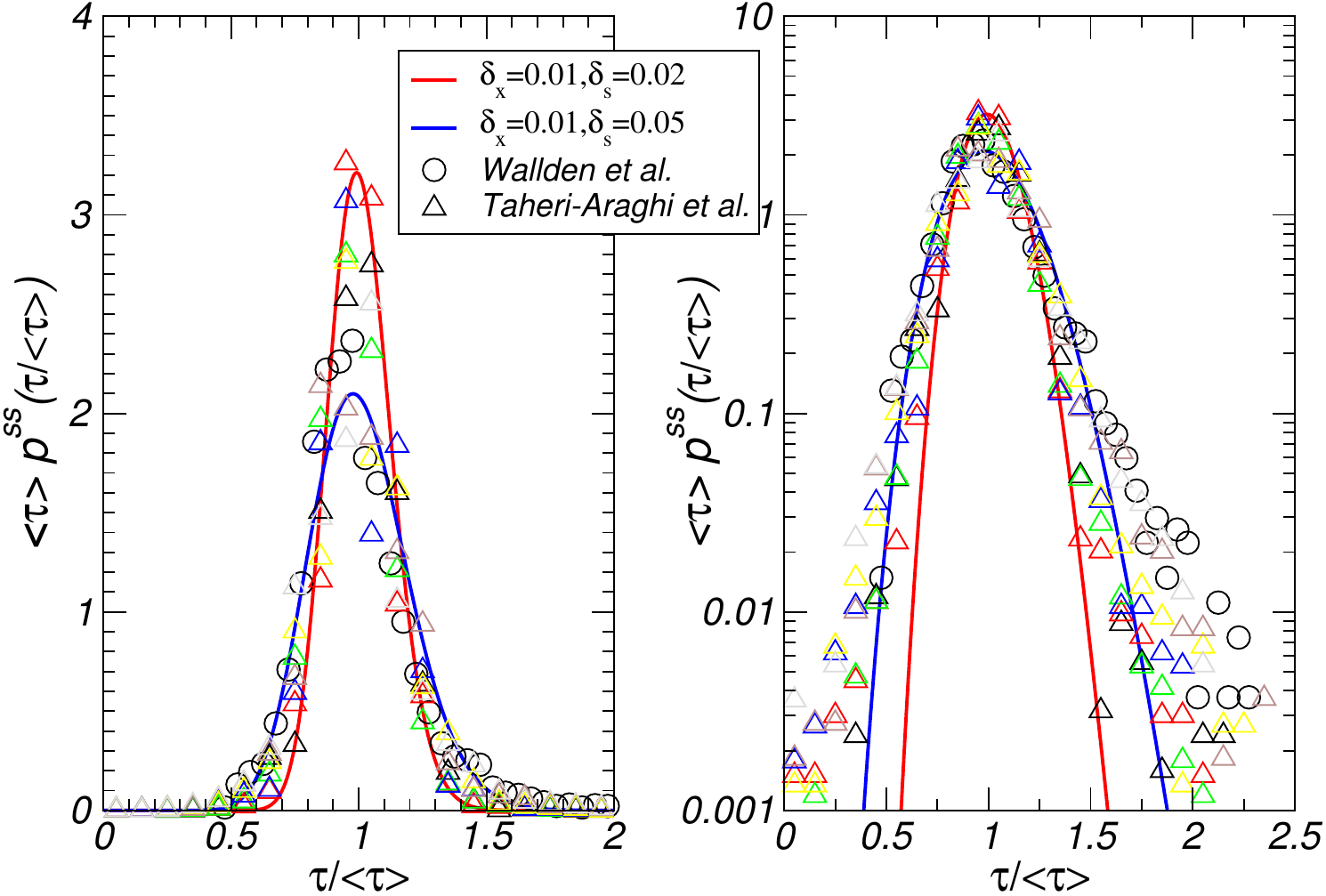}
\caption{\label{fig_taudist} The probability density function of the steady distribution of inter-division time  normalized by mean value. (Left) The lines show $\left<\tau\right>p^{ss}(\tau/\left<\tau\right>)$ for two typical noise levels. The symbols show the experimental data from \cite{suckjoon15,elf16}. (Right) Same as left, but in log-scale. 
}
\end{figure}

Following Eq.(\ref{eq_dldmatrix}), the distribution of division size evolves over generations according to $\hat{G}$. Noting the conditional probability $G\left(x^{(i+1)}_d\vert x^{(i)}_d\right)$ is normalized for any $x^{(i)}_d$, we see in the discrete version $\sum_m G_{mn}=1$ for any $n$. In other words, $\hat{G}$ is a Markov matrix. The distribution hence should evolve towards a steady one as the eigenvector of $\hat{G}$ by
\begin{equation}
P_x^{ss}=\hat{G}P_x^{ss}. 
\end{equation}
Provided Eq.(\ref{eq_das}) and $G_{mn}=P(\Delta=x_m-x_n/2)$, the eigen-problem can be solved in principle.  
With $P_x^{ss}$ as input, the steady distribution of inter-division time can be further evaluated by
\begin{equation}
P^{ss}(\tau)=\int dx\int dx'\; G(x^{(i+1)}_d=x,\tau\vert x^{(i)}_d=x')P^{ss}(x_d=x'). 
\end{equation}
For the above evaluation, the full Green's function, $G(x^{(i+1)}_d,\tau\vert x^{(i)}_d)$, is required. Since it is a joint distribution of $x_d^{(i+1)}$ and $\tau$ conditioned by $x_d^{(i)}$, one can write it as
\begin{equation}
\tilde{G}(x^{(i+1)}_d,\tau\vert x^{(i)}_d)=P(\tau\vert x^{(i+1)}_d,x^{(i)}_d )G(x^{(i+1)}_d\vert x^{(i)}_d),
\end{equation}
where $P(\tau\vert x^{(i+1)}_d,x^{(i)}_d )$ is the inter-division time distribution of a cell with known birth size $x_b=x^{(i)}_d/2$ and division size $x_d=x^{(i+1)}_d$. It may be estimated in the subordination picture of random walk, under the assumption that the waiting time for a jump from $(x_i,s_j)$ follows the exponential distribution with the mean value $\left<\tau_i\right>=[(\alpha+\beta)x_i]^{-1}$. This analytical approach is however too complicated. We use the numerically obtained $\tilde{G}(x^{(i+1)}_d,\tau\vert x^{(i)}_d)$ instead. 
Fig.\ref{fig_xd} and Fig.\ref{fig_taudist} show the theoretical results along with those of the recent experiments. Given different levels of cell cycle noises, the theoretical results capture the features of the experimental data, except the heavier tails observed in experiments.

We would like to remark here that the distribution of division size and inter-division time, $P(x_d,\tau_d)$ can also be obtained in the exponentially growing batch culture by counting the size and age of the dividing cells in a given time interval.
In the model defined by Eq.(\ref{eq_dme}) and  Eq.(\ref{eq_dbc}), the distributions can be evaluated by the division current 
\begin{equation}
P(x_d=x_i,\tau_d=\tau)=C \beta x_i u_{i,n_0}(\tau), 
\end{equation}
where $C=\left[\sum_i \int d\tau\; \beta x_i u_{i,n_0}(\tau)\right]^{-1}$ is introduced for normalization. 
The marginal distributions can be then evaluated for $x_d$ and $\tau$
\begin{equation}
P(x_d=x_i)=\int d\tau\; P(x_d=x_i,\tau_d=\tau),
\end{equation}
and
\begin{equation}
P(\tau_d=\tau)=\sum_i P(x_d=x_i,\tau_d=\tau). 
\end{equation}
In the long time limit, the memory on the sizes of the first cells is lost. The steady distributions are hence achieved. It has been confirmed that the distributions obtained in this approach well agree with the above results from the inter-generation stochastic dynamics, which is not shown here. 

\section{Noises level in cell growth and division processes: estimated from experimental data}

\label{sec_noise}

The added size distributions shown in Fig.\ref{fig_gf} suggest significant variance of noise strengths in various growth conditions. In this section, we estimate the strengths of growth and division noises from the experimental data. 

The analytic expression given by Eq.(\ref{eq_das}) and Eq.(\ref{eq_das1}) provide the exact solution of the added size distribution, of which the PDF is written as
\begin{equation}
\label{eq_das2}
p(\Delta;\alpha,\beta)=\binom{\alpha\Delta+\beta}{\beta}\frac{\alpha^{\alpha\Delta+1}\beta^{\beta+1}}{(\alpha+\beta)^{\alpha\Delta+\beta+1}}.
\end{equation}
It is difficult to directly apply Eq.(\ref{eq_das2}) for parameter estimation, due to numerical disasters brought by the drastic changing factors. For an applicable approximation, the improved form of Stirling's formula may be adopted that
\begin{equation}
\ln n!\simeq n\ln n-n+\frac{1}{2}\ln\left[\left(2n+\frac{1}{3}\right)\pi\right].
\end{equation}
Applying the formula to Eq.(\ref{eq_das2}), it gives
\begin{eqnarray}
\label{eq_lnpdf}
\ln p(\Delta;\alpha,\beta)&\simeq&(\alpha\Delta+\beta)\ln(\alpha\Delta+\beta)-\alpha\Delta\ln(\alpha\Delta)+(\alpha\Delta+1)\ln \frac{\alpha}{\alpha+\beta}\nonumber\\
& &+\beta\ln\frac{\beta}{\alpha+\beta}+\frac{1}{2}\ln\frac{\alpha\Delta+\beta+1/6}{\alpha\Delta+1/6}-\beta-\ln\Gamma(\beta).
\end{eqnarray}
This logarithmic PDF can be easily employed for the maximum likelihood estimation (MLE) for the parameters. Handling the experimental data, we would also note that the added size $\Delta$ in the above expression is the rescaled one $\Delta=\Delta_{\text{exp}}/(\gamma m_0)$, while the scaling factor $\Delta_0=\gamma m_0$ varies for experiments. Take it into account, the added size recorded by experiments, $\Delta_{\text{exp}}$, are expected following
\begin{equation}
\label{eq_lnpdf1}
\ln p(\Delta_{\text{exp}}=x;\alpha,\beta,\Delta_0)=\ln p(\Delta=x/\Delta_0;\alpha,\beta)-\ln\Delta_0.
\end{equation}
For each experimental dataset $\{x_i\}$, the MLE algorithm estimates the three parameters $\alpha,\beta$ and $\Delta_0$ by maximizing the logarithmic likelihood 
\begin{equation}
\mathcal{L}\left(\{x_i\}\right)=\sum_i \ln p(\Delta_{\text{exp}}=x_i;\alpha,\beta,\Delta_0). 
\end{equation}
Due to the non-Gaussian distribution, the transcendental equation arises in the MLE problem, which is numerically solved here. 

\begin{figure}[tbp]
	\centering
	\subfloat[Glucose\cite{suckjoon15}]{\label{fig_mle_1}
	\includegraphics[width=.30\textwidth]{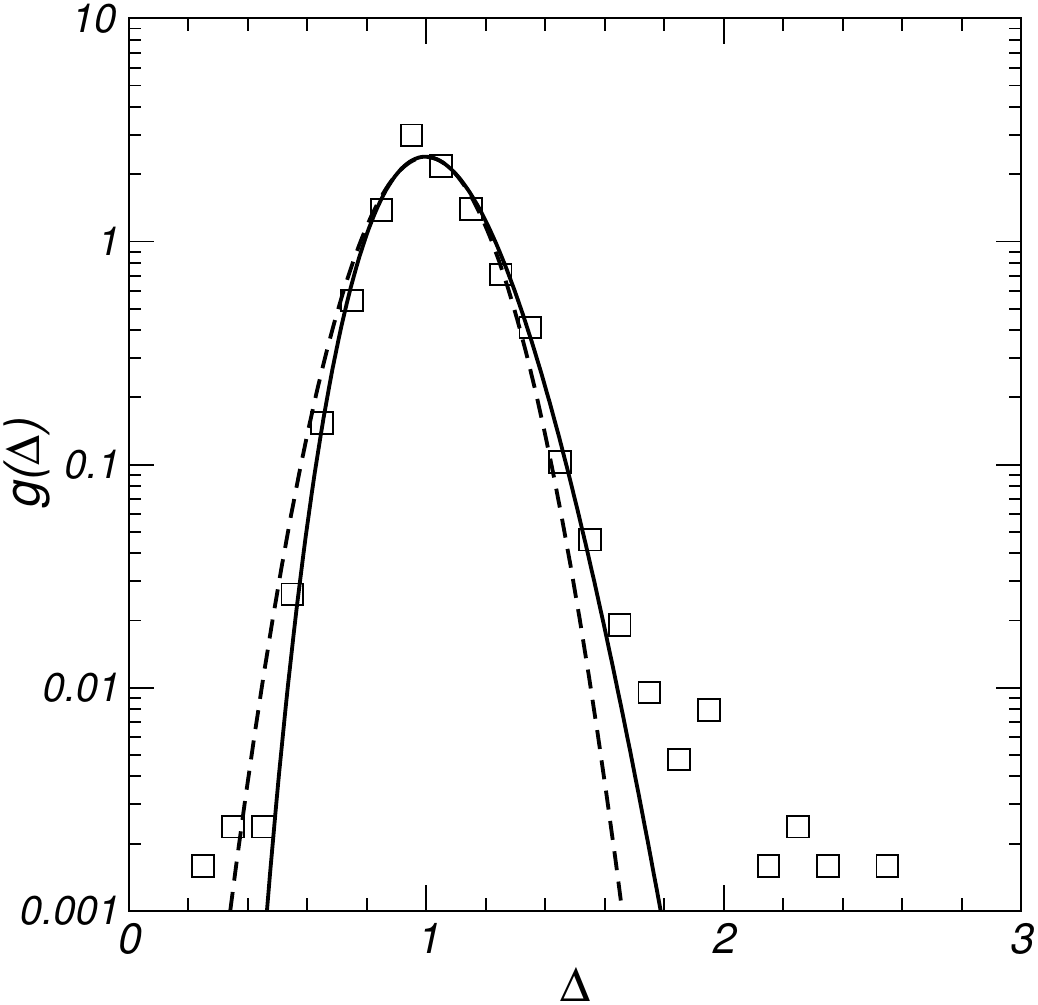}}\quad
	\subfloat[Glucose + 12 a.a\cite{suckjoon15}]{\label{fig_mle_2}
	\includegraphics[width=.30\textwidth]{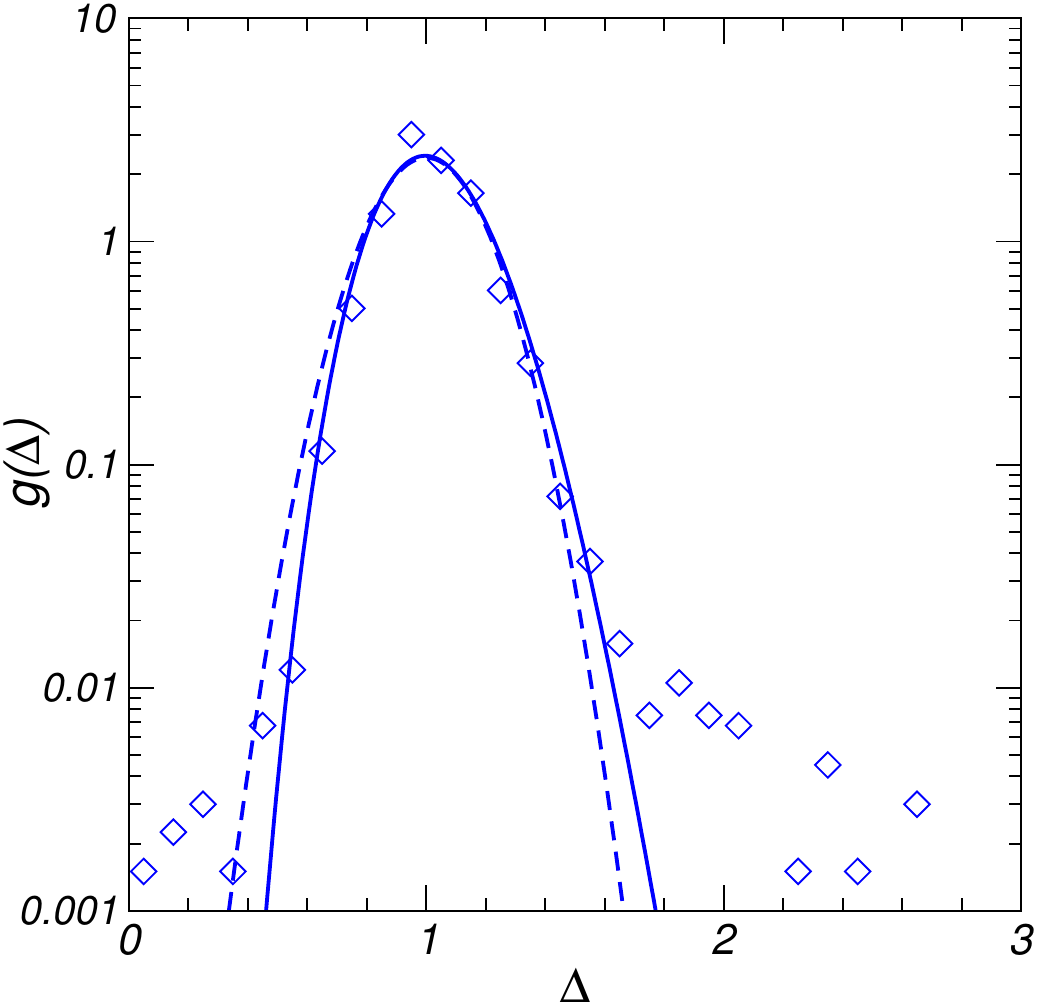}}\quad
	\subfloat[Glycerol\cite{suckjoon15}]{\label{fig_mle_3}
	\includegraphics[width=.30\textwidth]{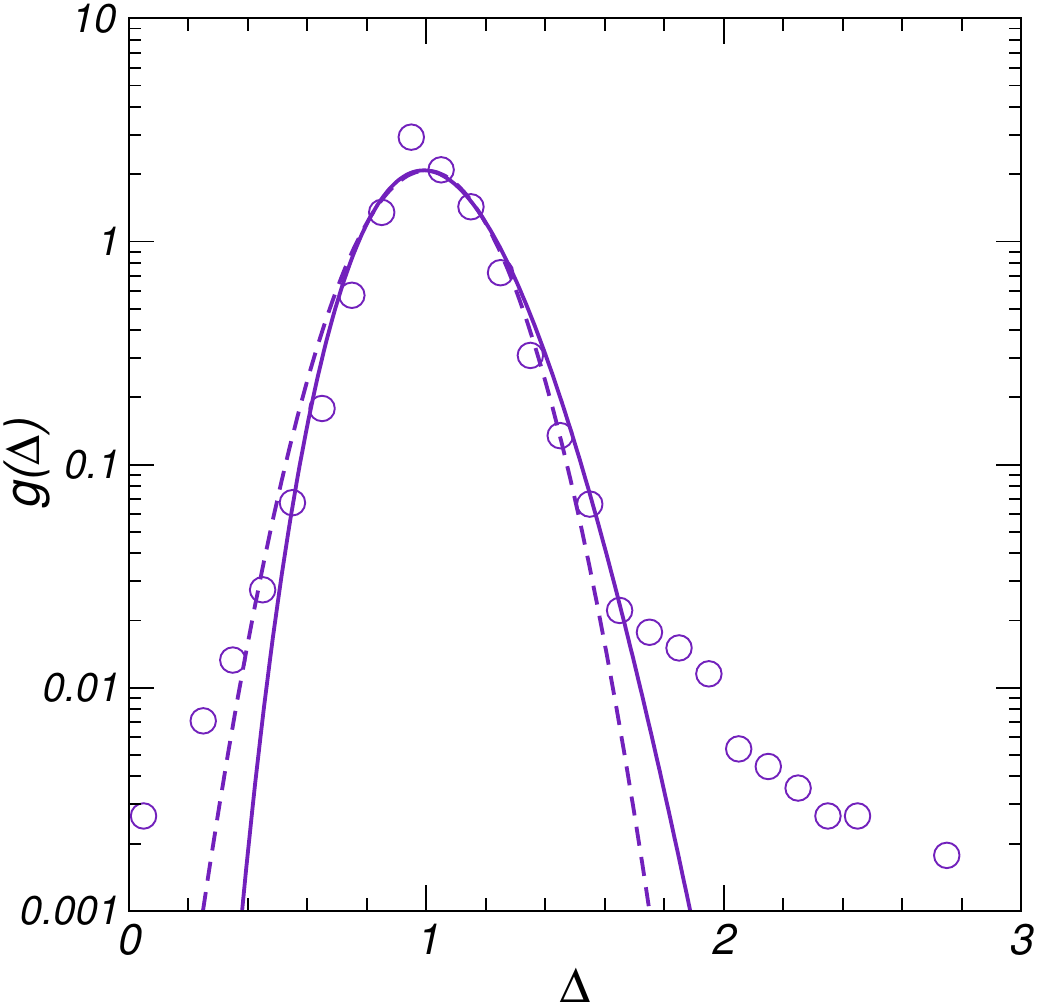}}\\
	\subfloat[Sorbitol\cite{suckjoon15}]{\label{fig_mle_4}
	\includegraphics[width=.30\textwidth]{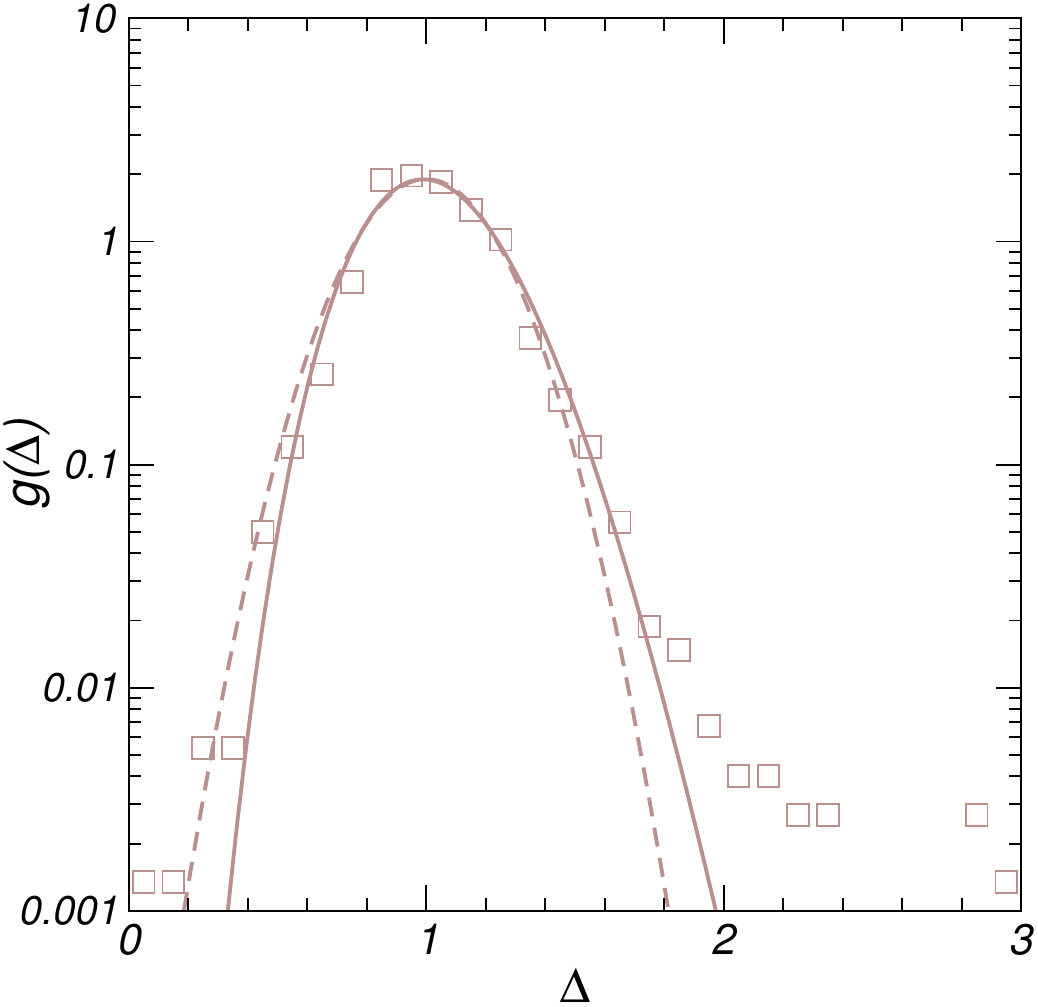}}\quad
	\subfloat[Fast growth condition\cite{elf16}]{\label{fig_mle_5}
	\includegraphics[width=.30\textwidth]{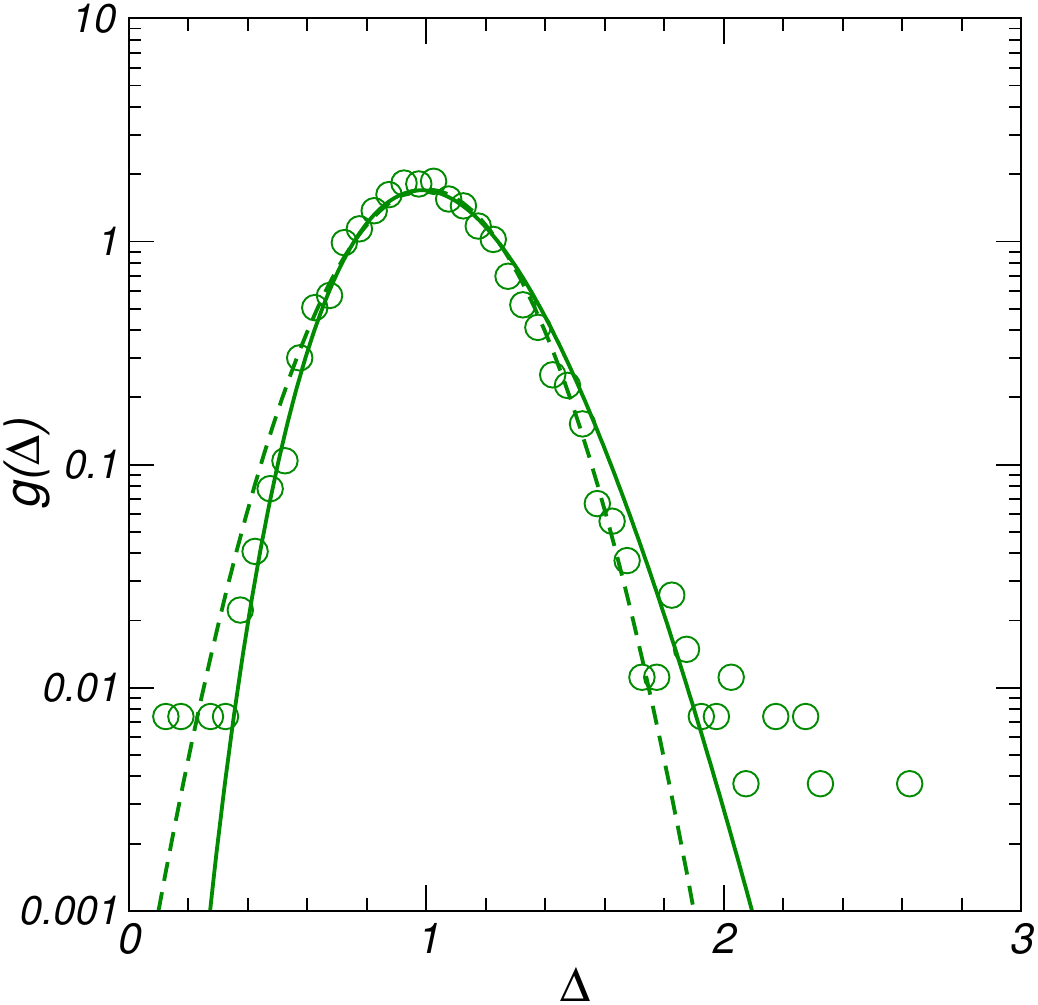}}\\
	\subfloat[Glucose + 6 a.a\cite{suckjoon15}]{\label{fig_mle_6}
	\includegraphics[width=.30\textwidth]{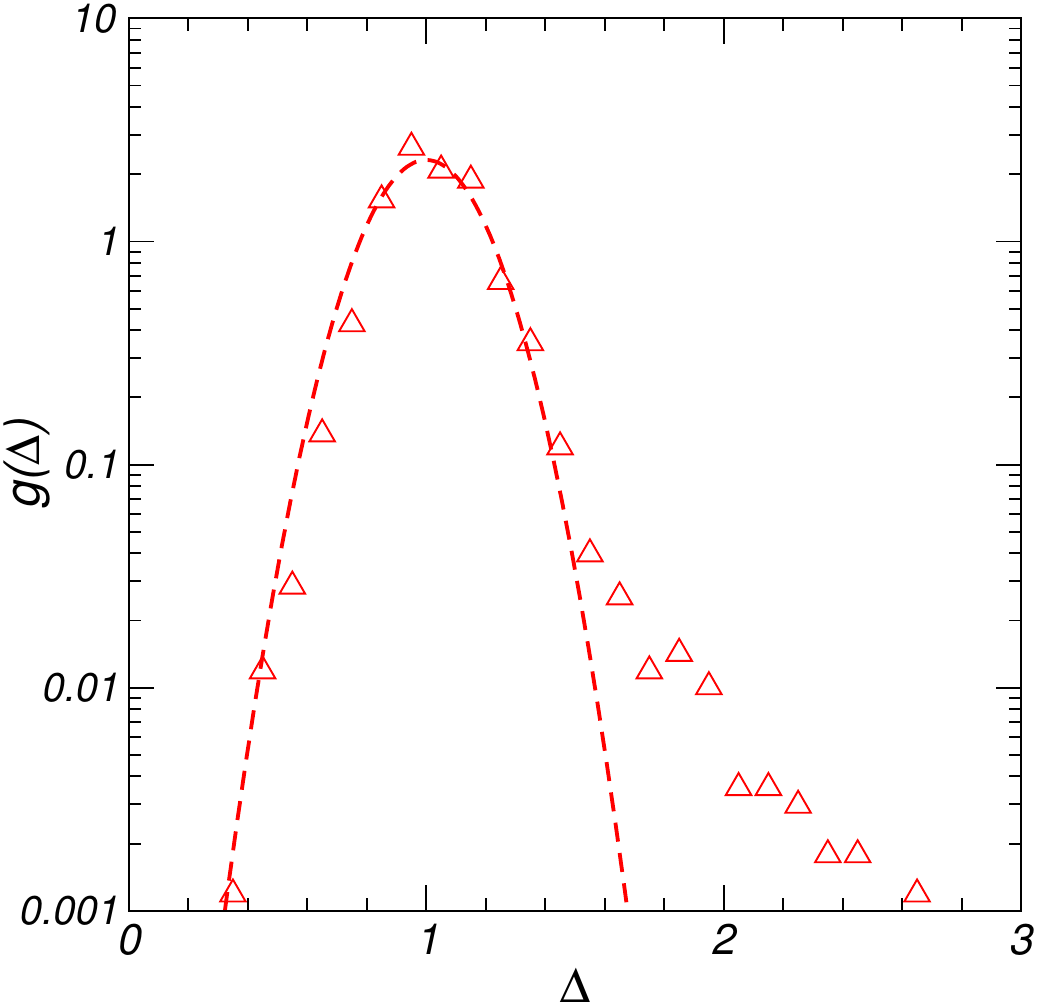}}\quad
		\subfloat[Synthetic rich\cite{suckjoon15}]{\label{fig_mle_7}
	\includegraphics[width=.30\textwidth]{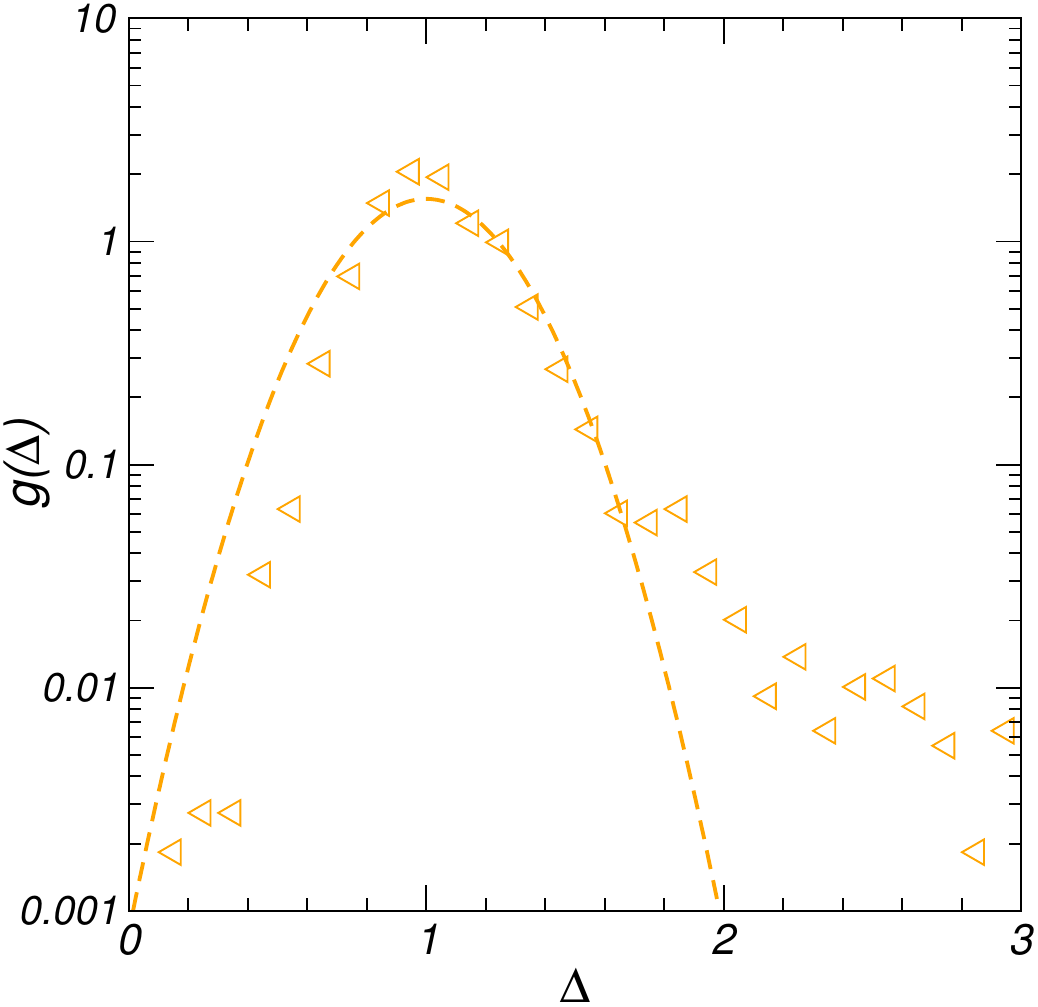}}\quad
	\subfloat[TSB\cite{suckjoon15}]{\label{fig_mle_8}
	\includegraphics[width=.30\textwidth]{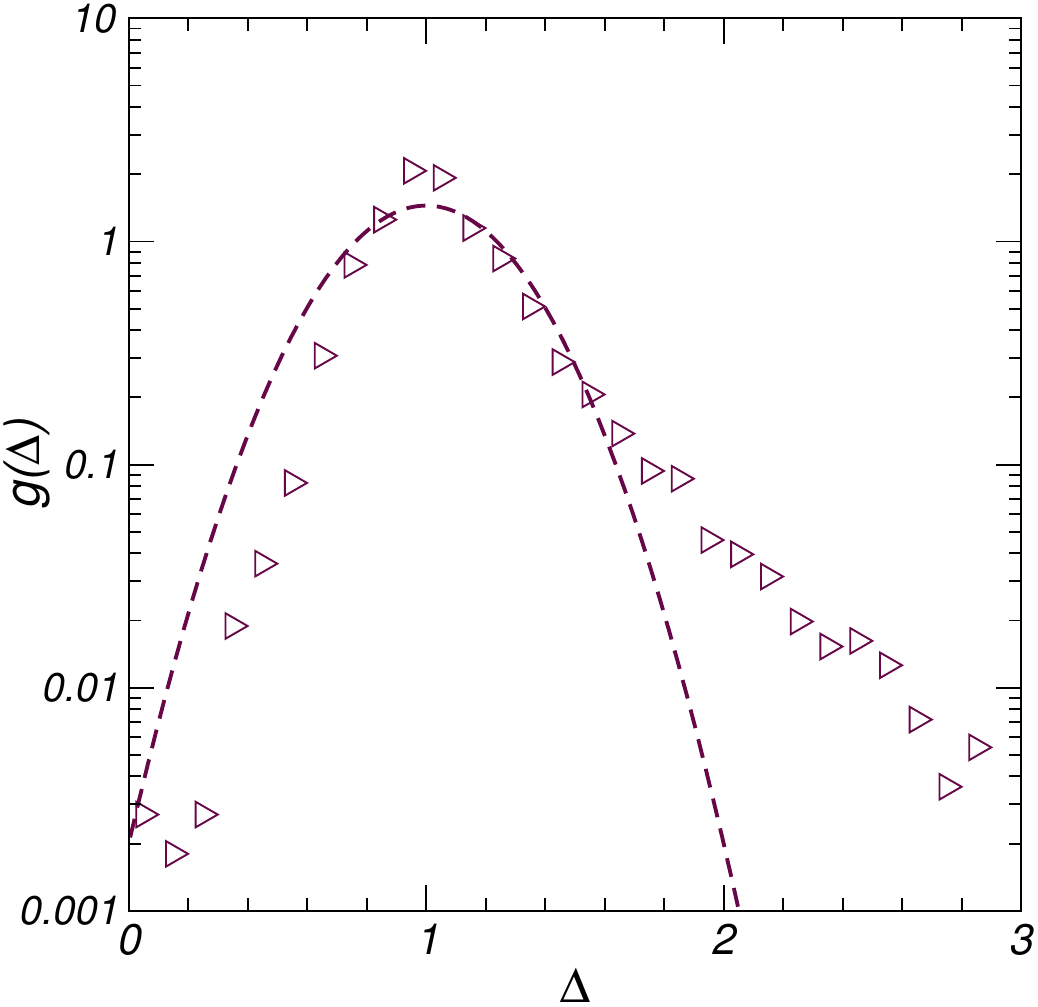}}\\
	\caption{\label{fig_mle}The PDF of added size estimated from the experimental data for various growth conditions. (Symbols) The experimental data.(Solid lines) The PDF estimated according to Eq.(\ref{eq_das2}). (Dashed lines) The PDF estimated according to the Gaussian distribution. In (f-h), the MLE according to Eq.(\ref{eq_das2}) is not stable due to the heavy tail. }
\end{figure}

\begin{table}
\centering
\caption{\label{tab_mle}The parameters estimated for experimental data of various growth conditions. The growth conditions with no stable MLE results are presented by --. The details of the stains and growth conditions can be found in the cited papers. } 
\begin{indented}
\lineup
\item[]\begin{tabular}{@{}*{7}{l}}
\br                              
\0\0Growth condition&$\alpha$&$\beta$&$\Delta_0$\cr 
\mr
Glucose \cite{suckjoon15}& $186.98$  & $44.92$ & $ 2.15$\cr
Glucose + 12 a.a. \cite{suckjoon15} & $129.27$  & $51.69$ & $ 2.88$\cr 
Glucose + 6 a.a. \cite{suckjoon15} & -- & --&-- \cr 
Glycerol \cite{suckjoon15} & $58.28$ & $51.34$ & $2.02$ \cr 
Sorbitol \cite{suckjoon15} & $44.42$ & $46.56$ & $2.10$\cr 
Synthetic Rich \cite{suckjoon15} & -- & -- & --\cr 
TSB \cite{suckjoon15} & -- & -- & --\cr 
Fast growth condition\cite{elf16} & $33.49$ & $39.69$ & $2.66$\cr 
\br
\end{tabular}
\end{indented}
\end{table}

We performed MLE for the seven growth conditions in \cite{suckjoon15} and the fast growth condition in \cite{elf16}. As shown in Table \ref{tab_mle}, the parameters have been stably estimated for five conditions out of eight, while MLE is not stable for the other three. All the data and the fitted distributions are shown in Fig.\ref{fig_mle}. The estimated parameters suggest that the noise of cell growth can vary for three folds for different nutrient conditions, while the division noises are more or less at the similar level. 

A heavy tail appears in the distribution of added size for all the experimental data sets, as shown in Fig.\ref{fig_mle}. One may already notice the heavy tail appears in every distribution of cell size for the considered experiments. (See Fig.\ref{fig_xdist}, Fig.\ref{fig_gf}, Fig.\ref{fig_xd}, and Fig.\ref{fig_mle}.)
A conjecture arises that it comes from the postponed cell division, the reason of which is still unknown in biology. It is because of the heavy tail, which significantly reshapes the distributions shown in Fig.\ref{fig_mle_6}-\ref{fig_mle_8}, the expressions from the current model (Eq.(\ref{eq_das2}), Eq.(\ref{eq_lnpdf}), and Eq.(\ref{eq_lnpdf1})) are not suitable to fit the data. While one can always add ingredients in model to fit the effects, our understanding can be truly improved only by careful experiment investigations on the rare events of the postponed division. The theory beyond the current accumulation model, based on novel biological insight, is required to embrace the strange tail. 



The Gaussian fitting is also attempted to fit the data, which is shown by the dashed lines in Fig.\ref{fig_mle}. Although the data are with considerable positive skewness, the Gaussian fitting with zero skewness still seems not too bad a choice except in the tail region. The following argument would help for better understanding. At the end of Sec.\ref{sec_as}, we have argued in the case of exponential growth with $x=x_b\exp(\gamma t)$, the Gaussian distributed inter-division time leads to the log-normal distributed division size, which is a good approximation of Eq.(\ref{eq_das2}). Assuming here the linear growth case with $x=x_b(1+\gamma t)$, the Gaussian distributed inter-division time would lead to the Gaussian distributed division size. Since the cell typically doubles the size over a cycle, $\lambda t\sim\ln2<1$ in the picture of exponential growth. For such $\lambda t$, the linear growth is a not-too-bad approximation of the exponential growth, and so for the Gaussian distribution as an approximation of the log-normal distribution and eventually of Eq.(\ref{eq_das2}). 

Although the non-Gaussian and Gaussian distributions of added size look similar, it will be shown in the next section the different features of the cell size control mechanism, with or without energy consumption. 

\section{Non-equilibrium nature of the cell size control dynamics}
\label{sec_ne}

When talking about the fluctuating yet stationary distributed cell size, one often uses the term {\it ``homeostasis''}. However, the precise meaning is not clear. In the view of statistical mechanics, one could go further to explicitly demonstrate the {\it non-equilibrium} nature of the cell size control mechanism, as shown in this section. 


Let us start from the master equation on the division sizes over generations, Eq.(\ref{eq_dldme}), and re-organize it as 
\begin{equation}
P^{(i+1)}_m-P^{(i)}_m=\sum_n J_{mn} -\sum_n J_{nm}, 
\end{equation}
where $J_{mn}=G_{mn}P^{(i)}_m$ is the transition current from $x_d^{(i)}=x_m$ to $x_d^{(i+1)}=x_n$. The two terms of the right-hand side are the inward current $J^{\text{in}}_m$ and the outward current $J^{\text{out}}_m$. When the balance between the inward and outward currents is achieved for all the $m$, i.e. $\sum_n G_{mn}P^{ss}_n =\sum_n G_{nm}P^{ss}_m$, the distribution becomes stationary over generations. This is the general balance condition for the steady state. As a special steady state, the equilibrium state requires a stricter constraint that the balance is achieved between any pair of states,  i.e. $G_{mn}P^{\text{eq}}_n=G_{nm}P^{\text{eq}}_m$, which is called the detailed balance condition.  For more background on non-equilibrium statistical physics, one can see e.g. \cite{seifert12}.

\begin{figure*}[tbp]
\centering

\subfloat{
\includegraphics[width=.4\textwidth]{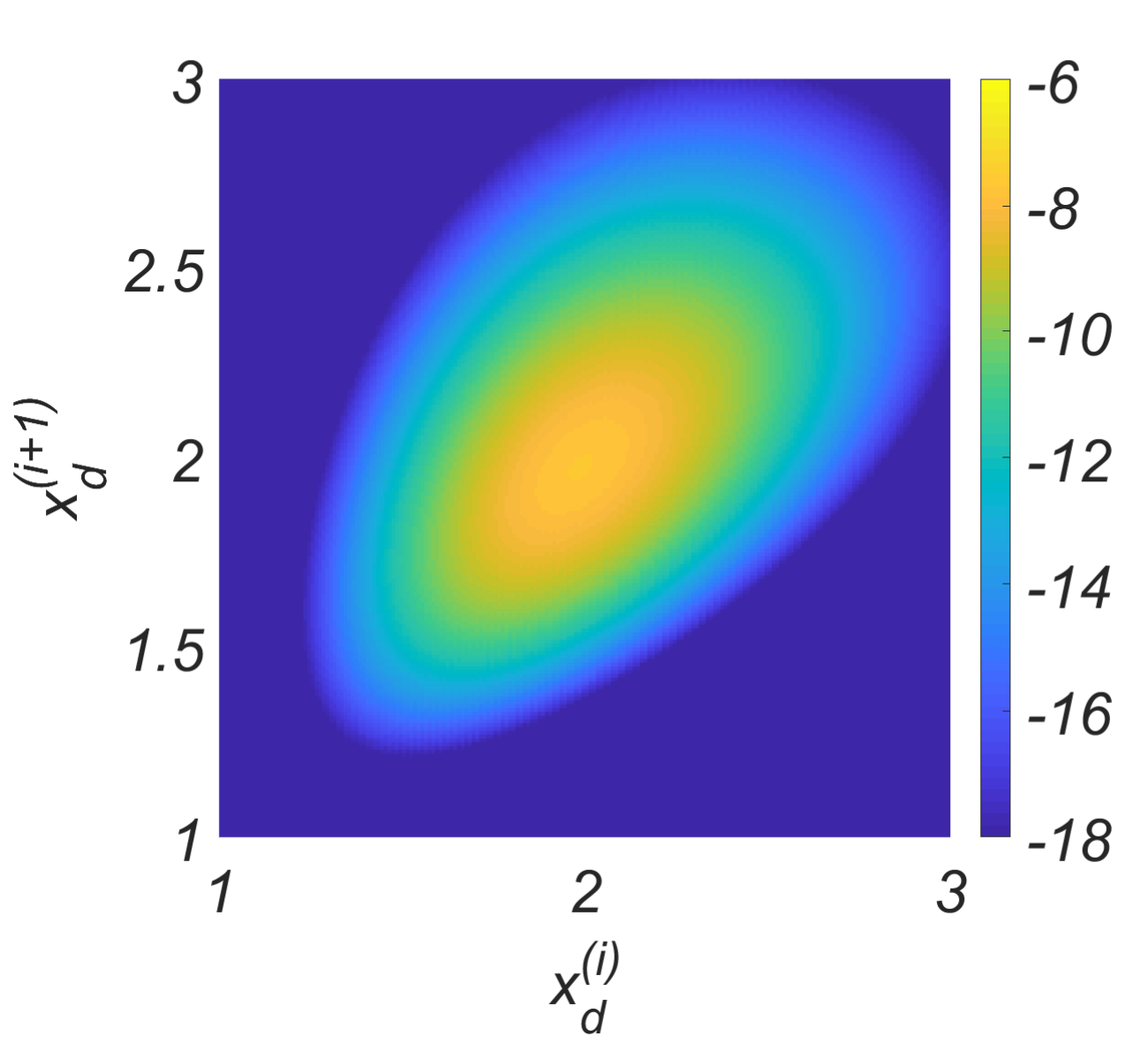}
}
\subfloat{
\includegraphics[width=.4\textwidth]{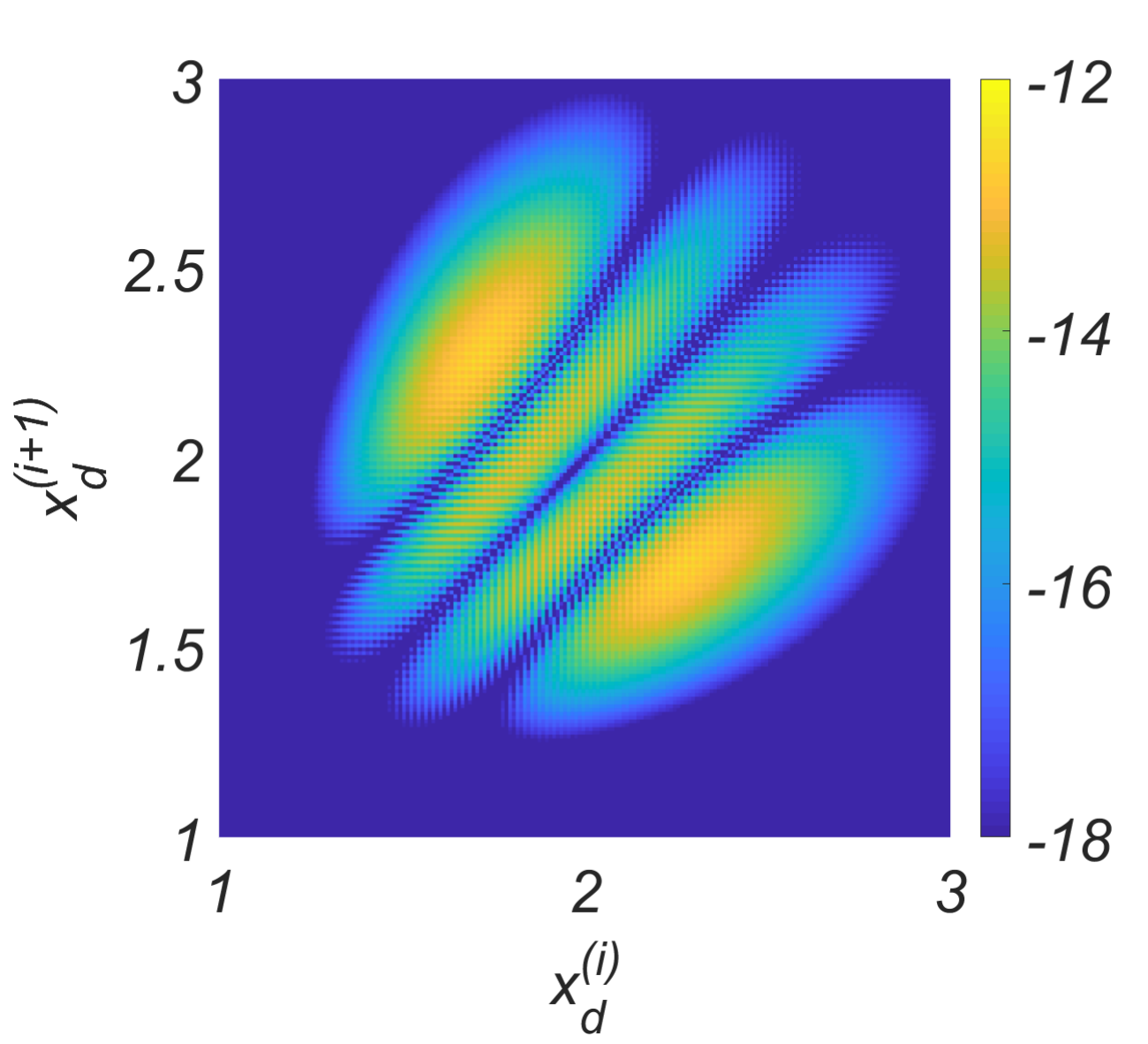}
}
\caption{\label{fig_epm} The heatmaps for the transition currents and entropy production in the steady state for $\delta_x=0.01$ and $\delta_s=0.02$. The colors are assigned according to the logarithm values. (Left) The transition currents $J^{ss}_{mn}$ between a pair of successive division sizes $x_d^{(i)}=x_m$ and $x_d^{(i+1)}=x_n$. (Right) The entropy production defined as $\Delta S_{mn}=\left(J^{ss}_{mn}-J^{ss}_{nm}\right)\left(\ln J^{ss}_{mn}-\ln J^{ss}_{nm}\right)/2$. One may note $\Delta S_{mn}=\Delta S_{nm}$ due to the definition.}
\end{figure*}

Equilibrium is usually assumed for the steady state of a closed physical system since it is the only state in which no entropy production (or energy dissipation) arises and hence requires no housekeeping energy. It is, however, not necessary for bacterial cells to satisfy this constraint. In contrast, a common strategy of living beings is to keep the system in non-equilibrium steady state (NESS) to best implement important biological functions at a cost of energy consumption\cite{lan12,lan16}. 
We notice that in this model of cell cycle, the transition current $J^{ss}_{mn}$ from $x_d^{i}=x_m$ to $x_d^{i+1}=x_n$ is asymmetric for $m$ and $n$, as shown in Fig.\ref{fig_epm}(left). The detailed balance between the division sizes is substantially broken. 
The non-equilibrium feature can be quantitatively measured by the entropy production\cite{seifert12} for each generation as
\begin{equation}
\Delta S=\sum_{m<n}\left(J^{ss}_{mn}-J^{ss}_{nm}\right)\left(\ln J^{ss}_{mn}-\ln J^{ss}_{nm}\right).
\end{equation}
The heatmap of $\Delta S_{mn}=\left(J^{ss}_{mn}-J^{ss}_{nm}\right)\left(\ln J^{ss}_{mn}-\ln J^{ss}_{nm}\right)/2$  is shown in Fig.\ref{fig_epm}(right). The non-trivial pattern reveals the structure of energy dissipation in the cell control process. 

Tracking the origin of the non-equilibrium feature, we find it is a consequence of the non-Gaussian noise on the added size. Assuming the Gaussian distributed $P(\Delta)$ in Eq.(\ref{eq_dld}) instead, the detailed balance would be recovered, which leads to an equilibrium scheme of cell size control. It is however not the case observed in experiments, where $P(\Delta)$ are reported with significant skewness. It is interesting to notice that Thomas find in his theoretical investigation on the population dynamics\cite{thomas18} that the non-Gaussian noise with the positive skewness, which is the current case, would decrease the cell-to-cell size variability in the population level. A connection between the energy consumption in the cell size control and its biological utility is indicated, which requires further investigation.

\section{Discussion}
\label{sec_disc}

\subsection{Historical remarks on the studies of cell size distribution of an exponentially growing population}
\label{sec_hr}
The bacterial cell cycle has been intensively investigated in the 1950s and 1960s by microbiologists in quite a quantitative way. A recent review has been given from the viewpoint of physicists\cite{suckjoon18}. Here we make short historical remarks on the achievement of cell size statistics. 
Since Schaechter, Maal\o{}e and Kjeldgaard's precise measurement on the bacterial cell sizes in 1958\cite{schaechter58}, the mechanism behind the well-controlled cell size was one of the most attractive topics in this field. In the 1960s, the Coulter counter was employed for automatic measurement of the size of cells in a culture, which drew theoretical interests to estimate the distribution. In 1961, Maclean and Munson discovered the "inverse square law of cell size distribution", saying $p(x)\sim x^{-2}$ for $x$ ranging from the minimum (new-born) cell size and the maximum (ready-for-division) cell size\cite{maclean61}, under the assumptions of exponential cell mass growth and perfect even-division with sharp distributed division sizes. It has been a well-known law in biology textbooks. Collins and Richmond\cite{collins62} and Koch\cite{koch62} independently obtained the general relation between the distribution of division size and the distribution of cell size of the whole population. Assuming normal distributed division size, Koch obtained the cell size distribution of the whole population\cite{koch66}. It suggested that the distribution was in general composed of the inverse-square peak blured by the fluctuation of division size. The shape of the distribution would be rather universal that the one obtained by Koch (Fig.1 and Fig.3 in \cite{koch66}) is quite similar to that in the current work (Fig.\ref{fig_xdist}), while the division size distributions are assumed in different forms. Noting the brilliant works on the cell size statistics, Bell, Anderson and their colleague from the physics community soon formulated the theory in the language of master equation\cite{bell67,bell67-2,bell68,bell69,bell71}, which led to the modern theoretical investigations. 

\subsection{Connections to other models}
\label{sec_comp}

In this work, bacterial cell cycle is investigated by the model of stochastic cell growth and adder indicator accumulation as random walk on the two-dimensional lattice, combined with divisions at the threshold $s=1$ for the adder indicator. There have been other approaches to construct the growth and division dynamics. Here we give a review of the previous studies and explain the connections to this study. 

Taheri-Araghi {\it et al.} developed the master equation for specific cell size control mechanisms\cite{suckjoon15} with the inputs from modern biological experiments. In their work, the master equation was kept in continuous form, while the stochastic division was defined by the rate function of the cell size ('sizer') or the added size ('adder'). With the distribution of elongation rate and the distribution of division size measured in the experiment as inputs, the theoretical results fit the experimental data very well. 

Noting the bursty feature of protein synthesis\cite{xie06}, various models of discrete dynamics have been introduced\cite{biswas14prl,singh16,warsawpre19,nieto20,jia21}. Ghusinga {\it et al.} studied the deterministic continuous cell size growth with the stochastic bursty accumulation of the adder indicator\cite{singh16}. The analytic expression of the distributions were obtained in their study on the first-passage problem, where the added size is shown independent of the birth size. Modifying the accumulation rate, Nieto {\it et al.} generalized the above bursty accumulation model to encompass various cell size control mechanism\cite{nieto20}. In this track, Jia {\it et al.} has recently investigated the bursty accumulation model with additional noise in cell partition\cite{jia21}. The cell size distribution and the inter-division time distribution have been analytically evaluated in the context of mother machine experiments. J\k{e}drak {\it et al.} studied the model of stochastic protein bursts and stochastic partition during division by an exactly solvable generalized Fokker-Planck equation, where the division rate is set to constant\cite{warsawpre19}. Ibis-Biswas {\it et al.} performed experiments on {\it Caulobacter crescentus} cells, in which experiment the collapse of cell size and inter-division time distributions was observed\cite{biswas14pnas}. The stochastic Hinshelwood cycle (SHC) model was introduced to explain their observation\cite{biswas14prl}. The SHC model is basically a master equation on discrete states, of which the boundary condition for cell division can be set in various ways. Digging into the historical literature, one can see that Kendall has proposed a similar model by considering the cell cycle as a Markov process exactly in the language of master equation of a stochastic birth process\cite{kendall48}. In the current work, we renew the theory by the accumulation mechanism, which is supported by recent bacterial experiments\cite{suckjoon15,willis17,suckjoon19}. The collapse of the distributions in the adder case arises in the stochastic model. Via the explicitly obtained Green's function, we bridged the accumulation model to the inter-generation processes.  

The inter-generation stochastic dynamics is a phenomenological description of the propagation of division features, such as the birth size or the division size, in cell lineages. It is a convenient language to investigate the cell size control mechanism\cite{amir14,you15,marco17}. In the current study on the accumulation model, the steady distributions are evaluated and well confirmed by the mother machine experimental data. Noting the non-Gaussian distribution of added size, we further predict the non-equilibrium nature of the inter-generation stochastic dynamics and the cell size control mechanism. 

We note in the last that the stochastic cell growth can be also introduced by adding diffusion terms to the continuous master equation (Eq.(\ref{eq_cme})), which leads to a Fokker-Planck (FP) equation. The equivalent Langevin dynamics of the FP equation has been investigated by Pirjol {\it et al.} for cell growth\cite{biswas17}. Under Kramers-Moyal expansion, the processes defined by the FP equation and the discrete master equation are equivalent to the second order. The third moments are different, which affect the skewness of the added size distribution. Physical reasoning of the difference comes that the cell size does not shrink during growth following the discrete master equation, while it is possible in that defined by the FP equation with naive diffusion terms. The moments of stochastic growth defined by the current model and the Fokker-Planck equation are estimated in \ref{app_moments}. 

\subsection{Statistics in batch culture and in mother machine}

In this study, the batch culture experiments and mother machine experiments are discussed separately because they are different in statistics. In batch culture experiments, due to the difficulty to track single cells for a long time, the statistics are usually on snapshots of the population, or saying, on cross-sections of the expanding lineage tree. In mother machine experiments, one can track few single branches on the lineage tree for generations, but not for all the offspring. The differences recalls the similar case in statistical mechanics, where one can either take a snapshot of the system by recording the states of all the particles at a certain time (known as the Gibbs ensemble), or track the motion of few particles and take the averages along the trajectories. 

To solve the different statistics, different languages have been developed to describe the two observations. The inter-generation stochastic dynamics focuses on the feature propagating along a single branch and the master equation concerns distributions in the view of population dynamics. In the current study, we have translated the master equation description into the inter-generation stochastic dynamics. The reverse translation has been formulated by Thomas by studying the statistics on a Cayley tree\cite{thomas18}. It is still subtle how to bridge two types of experiments by assigning proper weights to division events\cite{hashimoto16,thomas18}, especially in the case of large noises. 

\subsection{The noises in the cell cylce}

It has been a great achievement to construct the dynamics of bacterial cell cycle with the growth rate as the most important parameter, since the amazing finding\cite{schaechter58} by Schaecther {\it et al.} To understand why the system with external inputs of multi-degrees of freedom can be characterized by the single parameter, recent studies conducted by Hwa group constructed the coarse-grained model of cellular processes\cite{scott10,you13,basan15,erickson17}, which reveal the hidden principle  on the connection between growth rates and the partition of ribosome. This principle is well confirmed {\it on the level of mean value} by experiments with perturbations on various resources to break the degeneracy. 


Looking at the distribution of the data, the variance of the noise strengths can, however, not be neglected. The noise strengths are included in this study as the additional parameters. It is shown that the strength of the growth noise can vary for three folds in  various nutrient conditions. A biological reasoning comes that cell growth involves most of the noise sources in cell life\cite{thomas18nc}, including the nutrient uptake, the energy metabolism, the synthesis of macromolecules, etc. The noise strength may hence significantly vary for different growth conditions due to all the fluctuations in the processes. In contrast, the division noises mainly come from the bursty synthesis of few kinds of division-related proteins, which may be less affected by the fluctuations of other cellular processes and hence can be kept at the similar level. 

Although the bacterial physiological behavior has already been examined for various temperatures on the level of mean value in the seminal study by Schaechter {\it et al.}\cite{schaechter58}, the modern experiments on this topic controls bacteria mostly via nutrients and antibiotics. While the recent study on neurons can construct a clear model on the temperature-dependence of neuronal activities\cite{dvir18}, it is currently still a question how to integrate the temperature effects into the stochastic model of cellular processes. As the temperature is often thought to affect the noise level of cell growth and other reactions, it would be interesting to investigate the cell cycle of bacteria in the mother machine in thermostat of various temperatures, to provide the cellular information on noises and help to understand the temperature effect on cell cycle control. Combining the proteome information from the modern mass spectrometry techniques\cite{schmidt16}, it will further offer the opportunity to understand the molecular mechanisms helping bacteria survive in the wild nature of perpetually varying temperature.

\subsection{Biological implication of the decoupling between the birth size and the added size}
\label{sec_bi}
It was observed in the experiment\cite{wagner14,suckjoon15} that the added size is decoupled from the birth size. 
As a consequence, the added size distribution collapses for different birth sizes under a given growth condition\cite{suckjoon15}. 
Assuming the accumulation mechanism, certain connections between the cell growth and the accumulation of the adder indicator are expected. In the picture of random walk on the two-dimensional space of cell size and indicator abundance (see Fig.{\ref{fig_rw}}), the decoupling requires the ratio of cell growth rate $k^{(x)}$ and indicator accumulation rate $k^{(s)}$ are independent of cell size. In other words, $r_w= k^{(x)}/k^{(s)}$ may vary as a function of $s$ or other variables of the cell cycle but is expect with no direct correlation with cell size. In the language of biology, the expression level of the indicator, i.e. the fraction of the indicator mRNA in the whole transcriptome, is expected not to be a direct function of cell size, although it may change for different growth conditions or vary over a cell cycle.

\section{Summary}
\label{sec_cc}

The revolutionary modern techniques, including microfluidics and the advanced live imaging, lead to
the emergence of high-quality data on cell cycle at the single-cell level. While the celebrated theory on bacterial physiology has successfully explained the experimental observations on the level of mean value, the distributions of data suggest different noise levels for different growth conditions. The features of the noises are hence necessary for a theory concerning the distributions, which leads to the stochastic model studied here. 

In this theoretical work, we investigate the stochastic accumulation model in the framework of master equation. The inter-generation Green's function is analytically estimated for the model, which bridges two kinds of statistics used in the batch-culture and the mother machine experiments. It is shown that not only the mean value but also the distribution of added size is independent of the birth size. The picture of subordinated random walk is introduced to understand its biological origin, which is relevant to the expression level of the adder indicator. The theoretical results are utilized to analyze experimental data, which suggests that the division noise is kept in the similar level for different growth conditions while the strength of growth noise can vary for three folds. 
In the last, the non-equilibrium nature of bacterial cell size homeostasis is predicted by evaluating the entropy production in the inter-generation stochastic dynamics. The biological utility of the energy consumption during cell size control calls for further investigation. 

\ack
We are very grateful to Ramon Grima and Hai Zheng for comments and discussions. This work is supported by National Natural Science Foundation of China (Grant No. 11705064, 11804111) to L.L., National Key R\&D Program of China (Grant No. 2018YFA0903400), National Natural Science Foundation of China (Grant No. 32071417), CAS Interdisciplinary Innovation Team (Grant No. JCTD-2019-16), Guangdong Provincial Key Laboratory of Synthetic Genomics (Grant No. 2019B030301006) to X.F., National Natural Science Foundation of China (Grant No. 11804355, 31770111), Guangdong Provincial Natural Science Foundation (Grant No. 2018A030310010), Shenzhen Grant (Grant No. JCYJ20170413153329565) to Y.B..

\appendix

\section{The distribution of added size}
\label{app_dist}

In this appendix, we exactly solve the one-generation process for the distribution of division size $x_d$ of a cell with birth size $x_b$, $P(x_d\vert x_b)$. The distribution of added size can be directly obtained as $P(\Delta\vert x_b)=P(x_d=\Delta+x_b\vert x_b)$. 

The growth of a cell with birth size $x_b=i_0/a$ is defined by Eq.(\ref{eq_dme}), rewritten here as
\begin{equation}
\label{eq_dmea}
\frac{\partial}{\partial t}u_{ij}=-\alpha\left(x_i u_{ij}-x_{i-1}u_{(i-1)j}\right)-\beta\left(x_i u_{ij}-x_i u_{i(j-1)}\right),
\end{equation}
with the initial condition $u_{ij}\vert_{t=0}=\delta_{i,i0}\delta_{j,0}$, where $\alpha$, $\beta$ are the inversed noise strength and $\delta_{mn}$ is the Kronecker delta. The age term is omitted since $\tau=t$. 
Cell divides when $s=j\delta_s>1$, which can be formulated by the absorbing boundary condition, $u_{i,n_0+1}=0$ with $n_0=\beta=1/\delta_s$. The distribution of division size can be determined as the marginal distribution of the joint distribution of first-passage event
\begin{equation}
P(x_d=x_i\vert x_b)=\int_0^{\infty}dt\; F(x_d=x_i, t\vert x_b), 
\end{equation}
where the first-passage distribution is equivalent to the current of division events
\begin{equation}
F(x_d=x_i, t\vert x_b)=\beta x_i u_{i,n_0}(t).
\end{equation}

For convenience, we define $\tilde{u}_{ij}=\int_0^{\infty}dt\; u_{ij}(t)$ and $\tilde{v}_{ij}=x_i\tilde{u}_{ij}$. 
The division size distribution can be simply written as $P(x_d=x_i\vert x_b)=b\tilde{v}_{i,n_0}$.
Integrating over $t$ for both sides of Eq.(\ref{eq_dmea}), we obtain
\begin{equation}
u_{ij}\vert^{t=\infty}_{t=0}=-\alpha\left(\tilde{v}_{ij}-\tilde{v}_{i-1,j}\right)-\beta\left(\tilde{v}_{ij}-\tilde{v}_{i,j-1}\right).
\end{equation}
Noting that $u_{ij}(t=\infty)=0$ and $u_{ij}\vert_{t=0}=\delta_{i,i0}\delta_{j,0}$, one can get
\begin{equation}
(\alpha+\beta)\tilde{v}_{ij}-\alpha\tilde{v}_{i-1,j}=\beta\tilde{v}_{i,j-1}+\delta_{i,i_0}\delta_{j,0}.
\end{equation}
It defines the dynamics from $j-1$ to $j$ for the vector $\overrightarrow{v}^{(j)}=(\tilde{v}_{1j}, \tilde{v}_{2j},\cdots,\tilde{v}_{ij},\cdots)^{\text{T}}$ by
\begin{equation}
\begin{cases}
	\hat{M}\overrightarrow{v}^{(j)} =\beta\overrightarrow{v}^{(j-1)}, &\text{for }j=1,2,\cdots, \\
	\hat{M}\overrightarrow{v}^{(j)}=\overrightarrow{w}, & \text{for }j=0,
\end{cases}
\end{equation}
where $M_{mn}=(\alpha+\beta)\delta_{mn}-\alpha\delta_{m,n+1}$ and $w_i=\delta_{i,i_0}$. 
The formal solution is 
\begin{equation}
\overrightarrow{v}^{(j)}=\beta^{j}[\hat{M}^{-1}]^{(j+1)}\overrightarrow{w}. 
\end{equation}
It is fortunate that $\hat{M}^{-1}$ can be explicitly obtained as
\begin{equation}
\begin{cases}
	[\hat{M}^{-1}]_{mn} =\alpha^{m-n}(\alpha+\beta)^{-(m-n+1)}, & m\ge n\\
	[\hat{M}^{-1}]_{mn}=0, & m<n.
\end{cases}
\end{equation}
Noting 
\begin{equation}
\sum_{p=n}^{m}\binom{m-p+q}{q}=\binom{m-n+q+1}{q+1}, 
\end{equation}
one can easily obtain
\begin{equation}
\begin{cases}
	[\hat{M}^{-q}]_{mn} =\binom{m-n+q-1}{q-1} \alpha^{m-n}(\alpha+\beta)^{-(m-n+q)}, & m\ge n\\
	[\hat{M}^{-q}]_{mn}=0, & m<n.
\end{cases}
\end{equation}
It immediately leads to the explicit solution
\begin{equation}
\tilde{v}_{ij}=\overrightarrow{v}^{(j)}_i=\beta^j[\hat{M}^{-(j+1)}]_{i,i_0}=\binom{i-i_0+j}{j}\frac{\alpha^{i-i_0}\beta^j}{(\alpha+\beta)^{i-i_0+j+1}}.
\end{equation}
Noting $x_i=i/\alpha$, $x_b=i_0/\alpha$ and $s_0=n_0/\beta=1$ and also $P(x_d=x_i\vert x_b)=\beta\tilde{v}_{i,n_0}$.
one can eventually obtain the added size distribution as
\begin{equation}
P(\Delta\vert x_b)=P(x_d=x_i+\Delta\vert x_b)=\binom{\alpha\Delta+\beta}{\beta}\frac{\alpha^{\alpha\Delta}\beta^{\beta+1}}{(\alpha+\beta)^{\alpha\Delta+\beta+1}}, 
\end{equation}
which is Eq.(\ref{eq_das}) in the main text. 
The probability density function can be get as
\begin{equation}
g(\Delta)=\alpha P(\Delta\vert x_b).
\end{equation}

\section{The moments and cumulants of the stochastic dynamics of cell growth: master equation and Fokker-Planck equation}
\label{app_moments}

In this appendix, we estimate the time course of moments and cumulants during cell growth. 

Let us recall the deterministic process defined by the continuous master equation,
\begin{equation}
\frac{\partial }{\partial t}u=-\frac{\partial }{\partial x}(x u)-\frac{\partial }{\partial s} (xu),
\end{equation}
of which the ageing term is neglected for the one-generation growth $\tau$ is alway equal to $t$. 
The stochasticity is introduced by the parameters $\delta_x$ and $\delta_s$ in the discrete master equation (ME)
\begin{equation}
\frac{\partial}{\partial t}u_{ij}=-\frac{1}{\delta_x}\left(x_i u_{ij}-x_{i-1}u_{(i-1)j}\right)-\frac{1}{\delta_s}\left(x_i u_{ij}-x_i u_{i(j-1)}\right). 
\end{equation}
For a cell with the given birth size $x=x_b$, the initial condition is written as $u_{ij}\vert_{t=0}=\delta_{i,i_0}\delta_{s,0}$, where the Kronecker delta is employed and $i_0=x_b/\delta_b$. 
The moment can be defined by
\begin{equation}
\left<x^m s^n\right>=\sum_{i,j}x_i^m s_j^n u_{ij}. 
\end{equation}
Applying the definition to the master equation, it gives
\begin{equation}
\frac{d}{d t}\left<x^m s^n\right>=-\frac{1}{\delta_x}\sum_{i,j}\left(x_i^{m+1} s_j^n-x_{i+1}^{m} x_{i}s_j^n \right)u_{ij}-\frac{1}{\delta_s}\sum_{i,j}\left(x_i^{m+1} s_j^n-x_i^{m+1} s_{j+1}^n\right)u_{ij}, 
\end{equation}
while the initial condition is $\left<x^m\right>\vert_{t=0}=x_b^m$ and $\left<x^m s^n\right>\vert_{t=0}=0$ for $n>0$.
Noting $x_i=i \delta_x$ and $s_j=j \delta_s$, it can be read that the time course of the first several moments are determined by the equations
\begin{eqnarray}
\frac{d }{dt}\left<x\right>&=&\left<x\right>,\nonumber\\
\frac{d }{dt}\left<x^2\right>&=&2 \left<x^2\right>+\delta_x\left<x\right>,\nonumber\\
\frac{d }{dt}\left<x^3\right>&=&3 \left<x^3\right>+3\delta_x\left<x^2\right>+\delta_x^2\left<x\right>, \nonumber\\
\frac{d }{dt}\left<s\right>&=&\left<x\right>,\nonumber\\
\frac{d }{dt}\left<xs\right>&=&\left<xs\right>+\left<x^2\right>,\nonumber\\
\frac{d }{dt}\left<s^2\right>&=&2\left<xs\right>+\delta_s\left<x\right>.\nonumber
\end{eqnarray}
The moments are solved from the above equations with the initial conditions, which give the cumulants of $x$
\begin{eqnarray}
\mu_{\text{ME}}&=&\left<x\right>=x_b e^t, \nonumber\\
\sigma^2_{\text{ME}}&=&\left<x^2\right>-\left<x\right>^2=\delta_x x_b e^t(e^t-1),\nonumber\\
\kappa^3_{\text{ME}}&=&\left<\left(x-\left<x\right>\right)^3\right>=\delta_x^2 x_b e^t (e^t-1)(2e^t-1),\nonumber
\end{eqnarray}
and also those for $s$
\begin{eqnarray}
\left<s\right>&=&x_b (e^t-1), \nonumber\\
\left<s^2\right>-\left<s\right>^2&=&\delta_sx_b (e^t-1)+\delta_x x_b(e^{2t}-2t e^{t}-1).\nonumber
\end{eqnarray}
The higher moments and cumulants can be obtained recursively, which are not shown here. 

In the Fokker-Planck(FP) approach, the equation is kept in the continuous form, while the diffusion terms are introduced as
\begin{equation}
\frac{\partial }{\partial t}u=-\frac{\partial }{\partial x}(x u)+\frac{\delta_x}{2}\frac{\partial^2}{\partial x^2}(xu) -\frac{\partial }{\partial s}(xu)+\frac{\delta_s}{2}\frac{\partial^2}{\partial s^2}(xu),
\end{equation}
The initial condition is $u\vert_{t=0}=\delta(x-x_b)\delta(s)$, where the Dirac delta function appears. 
The moment is defined in the continuous case as
\begin{equation}
\left<x^m s^n\right>=\int dx\int ds\;x^m s^n u(x,s). 
\end{equation}
Assuming $\lim_{x\rightarrow\infty}u(x,s)=0$ and $\left<x^{m}\vert s=0\right>=0$ for any $m>1$, one can again obtain the equations for the moments similar to the above ME case. The equations are the same until $m=3$, where
\begin{equation}
\frac{d }{d  t}\left<x^3\right>=3 \left<x^3\right>+3\delta_x\left<x^2\right>.\nonumber
\end{equation}
The cumulants are hence given as
\begin{eqnarray}
\mu_{\text{FP}}&=&x_b e^t=\mu_{\text{ME}}, \nonumber\\
\sigma^2_{\text{FP}}&=&\delta_x x_b e^t(e^t-1)=\sigma^2_{\text{ME}},\nonumber\\
\kappa^3_{\text{FP}}&=&\frac{3}{2}\delta_x^2 x_b(e^t-1)^2 e^t<\kappa^3_{\text{ME}}.\nonumber
\end{eqnarray}
One can see the two stochastic approaches are equivalent to the second order. It recalls us the fact that the Fokker-Planck equation is the second-order approximation of the master equation under Kramer-Moyal expansion\cite{riskenbook}. The difference arises for the higher moments, resulting in different skewness $\kappa^3$ of the size distribution. 

With the expression of the cumulants, one may try the Gram-Charlie series to obtain the analytic expression of $u(x,s)$\cite{sauer79}. It would however not help, due to the serious minus probability problem when the higher-order terms with $m\ge3$ are included, which is also a known feature of Kramer-Moyal expansion\cite{riskenbook}. 

We note that the Langevin dynamics corresponding to the above FP approach has been studied by Pirjol {\it et al.}\cite{biswas17}, where the same results on $\left<x\right>$ and $\left<x^2\right>$ were obtained. 

\section{The mean value, coefficient of variation, and skewness of the added size}
\label{app_noise}

\begin{figure}
	\centering
	\subfloat{\label{fig_cv}
	\includegraphics[width=.4\textwidth]{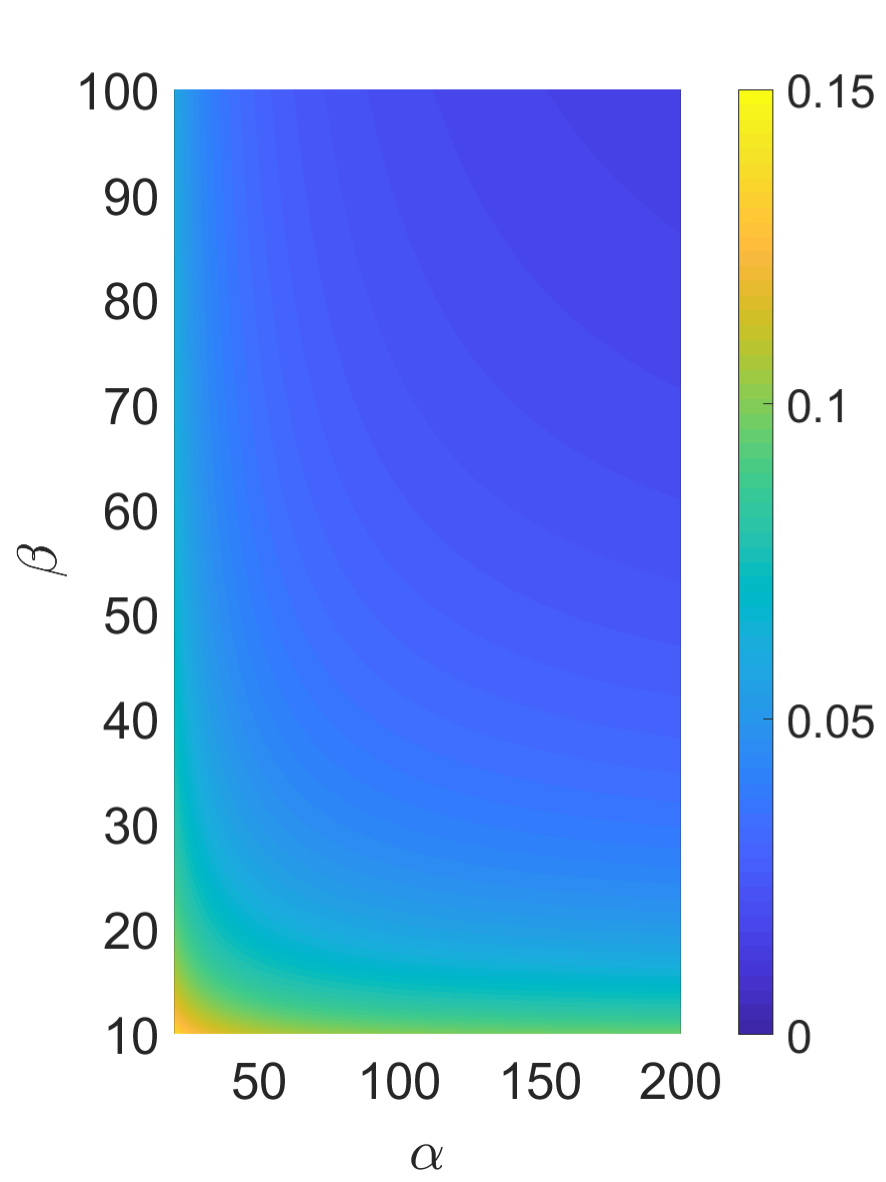}}
	\subfloat{\label{fig_skewness}
	\includegraphics[width=.4\textwidth]{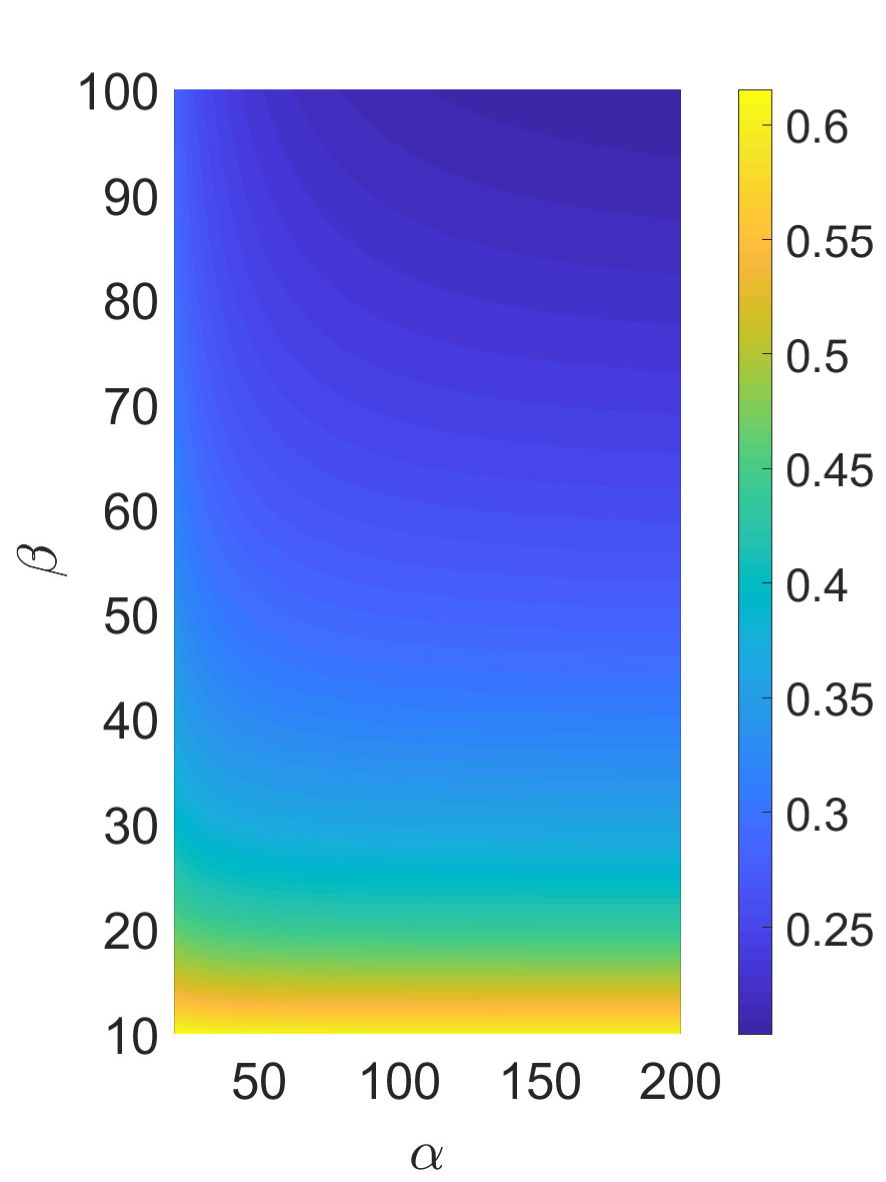}}
	\caption{\label{fig_cvskewness} The heatmap of the coefficient of variation (left) and skewness (right) for different noise levels. }
\end{figure}

In the deterministic accumulation model, the added size is determined by the growth rate $\gamma$ and the mean rescaled added size $m_0$ by $\Delta=\gamma m_0$. In present of noises, the mean value of the added size can be evaluated from the analytic expression as 
\begin{equation}
\left<\Delta\right>=\gamma m_0 \left(1+\frac{1}{\beta}\right).
\end{equation}
It slightly shifts from that by the deterministic model due to the noise of division, while the growth noise would not affect it. As estimated from experimental data, $\beta\sim50$, the correction to the mean value is not significant. The coefficient of variation (CV) and skewness can be further evaluated as
\begin{equation}
\text{CV}\equiv\frac{\left<\Delta-\left<\Delta\right>^2\right>}{\left<\Delta\right>^2}=\frac{\alpha+\beta}{\alpha+\alpha\beta}=\frac{1+\mu}{1+\beta},
\end{equation}
and
\begin{equation}
\text{Skewness}\equiv\frac{\left<\Delta-\left<\Delta\right>^3\right>}{\left<\Delta-\left<\Delta\right>^2\right>^{3/2}}=\frac{2\alpha+\beta}{\sqrt{\alpha(1+\beta)(\alpha+\beta)}}=\frac{2+\mu}{\sqrt{(1+\beta)(1+\mu)}},
\end{equation}
where the ratio of the inversed noise strengths, $\mu=\beta/\alpha$, is introduced. 
In the limit of the deterministic and continuous growth, $\alpha\rightarrow\infty$ and $\mu\rightarrow0$, 
$\text{CV}\rightarrow 1/(1+\beta)$ and $\text{Skewness}\rightarrow 2/\sqrt{1+\beta}$. 
The CV and Skewness for various $\alpha$ and $\beta$ are shown in Fig.\ref{fig_cvskewness}.

\section{The pseudocode for cell cycle simulation}
\label{app_pseudocode}

This appendix provides the pseudocode of cell cycle simulation by Gillespie's algorithm\cite{gillespie77}. To be noted, most of the results in this work are obtained by analytic or numerical approaches instead of simulation. 

\begin{algorithm}
\caption{Cell cycle simulation by Gillespie's algorithm}
\begin{algorithmic} 
\STATE $x\leftarrow x_b$
\STATE $s\leftarrow 0$
\STATE $\tau\leftarrow 0$
\STATE $t\leftarrow 0$
\STATE Update $k^{(s)}$ and $k^{(x)}$ according to $x$
\STATE
\WHILE{$t<t_{\text{max}}$}
\STATE
\STATE $t_0 \leftarrow 1/(k^{(s)}+k^{(x)})$			
\STATE Draw a random number $\delta t$ from exponential distribution with mean value $t_0$
\STATE $t\leftarrow t+\delta t$
\STATE $\tau\leftarrow \tau+\delta t$
\STATE
\STATE Draw a random number $jump$ uniformly distributed in $[0,1]$
\IF{$jump < k^{(s)}/(k^{(s)}+k^{(x)})$}
\STATE $s \leftarrow s+\delta_s$
\ELSE
\STATE $x \leftarrow x+\delta_x$
\STATE Update $k^{(s)}$ and $k^{(x)}$ according to $x$
\ENDIF
\STATE
\IF{$s\ge s_0$}
\STATE $s\leftarrow 0$
\STATE $x\leftarrow x/2$
\STATE $\tau\leftarrow 0$
\ENDIF
\STATE
\ENDWHILE
\end{algorithmic}
\end{algorithm}

\section*{References}
\bibliography{cellcycle}

\end{document}